\documentclass[twocolumn,showpacs,preprintnumbers,amsmath,amssymb,pra,floatfix]{revtex4-2}
\usepackage{color}

\usepackage{graphicx}% Include figure files
\usepackage{dcolumn}% Align table columns on decimal point
\usepackage{bm}% bold math
\usepackage{epstopdf}
\usepackage{bbold}
\usepackage{amsthm}
\usepackage{moresize}
\usepackage[colorlinks,linkcolor = blue,anchorcolor = blue, urlcolor = blue,citecolor = blue,pdfauthor=author ]{hyperref}

\newtheorem{theorem}{Theorem}
\newtheorem{definition}{Definition}
\newtheorem{lemma}{Lemma}
\newtheorem{corollary}{Corollary}

\newcommand{\iden}{\mathbb{1}}

\newcommand{\tr}{\mathrm{Tr}}
\newcommand{\operation}{\Lambda}
 
\newcommand{\W}{{\tiny W}}

\newcommand{\FO}{\operation^\mathrm{F}}

\newcommand{\mmtpd}{\mathrm{MMTP}^{(d)}}

\newcommand{\mmtpt}{\mathrm{MMTP}^{(2)}}

\newcommand{\tp}{\mathrm{TP}}
\newcommand{\mtp}{\mathrm{MTP}}
\newcommand{\fo}{\mathrm{F}}
\newcommand{\etp}{\mathrm{ETP}}
\newcommand{\mE}{\mathcal{E}}
\newcommand{\mT}{\mathcal{T}}

\makeatletter

\newcommand{\Rmnum}[1]{\expandafter\@slowromancap\romannumeral #1@}
\makeatother

\begin{document}

\preprint{}

\title{Cooling and work extraction under memory-assisted Markovian thermal processes}% Force line breaks with \\
\author{Yuancheng Liu}

\author{Xueyuan Hu}
\email{xyhu@sdu.edu.cn}
\affiliation{School of Information Science and Engineering, Shandong University, Qingdao 266237, China}%Lines break automatically or can be forced with \\

\date{\today}% It is always \today, today,
             %  but any date may be explicitly specified

\begin{abstract}
We investigate the limits on cooling and work extraction via Markovian thermal processes assisted by a finite-dimensional memory. Here the memory is a $d$-dimensional quantum system with trivial Hamiltonian and initially in a maximally mixed state.
For cooling a qubit system, we consider two paradigms, cooling under coherent control and cooling under incoherent control. For both paradigms, we derive the optimal ground-state populations under the set of general thermal processes (TP) and the set of Markovian thermal processes (MTP), and we further propose memory-assisted protocols, which bridge the gap between the performances of TP and MTP. For the task of work extraction, we prove that when the target system is a qubit in the excited state, the minimum extraction error achieved by TP can be approximated by Markovian thermal processes assisted by a large enough memory. Our results can bridge the performances of TP and MTP in thermodynamic tasks including cooling and work extraction.
\end{abstract}

%\pacs{05.30.-d, 03.65.Ta, 03.67.-a}% PACS, the Physics and Astronomy
                             % Classification Scheme.
%\keywords{Suggested keywords}%Use showkeys class option if keyword
                              %display desired
\maketitle

\section{Introduction}
Laws of thermodynamics pose essential restrictions on processes such as cooling and work extraction.
% The problem of work extraction from a non-equilibrium state is origninated from Szilard's discussion of maxwell's demon, where the amount of extracted work is related to the knowledge of the target system in the asymptotic regime.
Ever since Szilard's discussion of Maxwell's demon, where the amount of extracted work is related to the knowledge of the target system, the information-theoretic approach has been employed to investigate thermodynamic processes \cite{RevModPhys.81.1}. In particular, proposals for quantum engines have been proposed \cite{PhysRevLett.123.250606,PhysRevLett.111.230402,PhysRevX.7.021003,PhysRevResearch.4.033233} and experimentally demonstrated \cite{PhysRevLett.126.080603,PhysRevLett.121.030604,PhysRevLett.121.180601,WOS:000563802000001}. Further, quantum gates are employed to design schemes for cooling, where the entropy of the target system is transferred to an auxiliary system, which can then release the entropy to a heat bath. Such scheme of cooling is called heat bath algorithmic cooling \cite{PNAS_HBAC,PhysRevLett.94.120501,PhysRevLett.114.100404,PhysRevLett.123.170605}.

Recently, great progresses have been made in resource-theoretic approaches to quantum thermodynamics \cite{PhysRevLett.111.250404,Lostaglio_2019}. In particular, conditions on state transformations under the set of thermal processes (TP) are established \cite{Horodecki2013thermalmajor}. Based on these conditions, bounds on work extraction are derived \cite{Horodecki2013thermalmajor}. Also, limitations on cooling in different regimes are derived \cite{srep05192,PhysRevE.84.041109,PhysRevX.7.041033,PRXQuantum.4.010332}. Further, the heat-bath algorithmic cooling is extended to involve general thermal processes, such that the ground-state population of the target system goes to 1 exponentially fast in the number of rounds \cite{Alhambra2019heatbathalgorithmic}.

 % Great progress in resource-theoretic (quantum information theoretic) approach.\\
 % quantum engine.\\
 % Algorithmic cooling.\\

 When the set of thermal processes is restricted by the Markovian condition, which means that the evolution of states under such processes can be described by a master equation with the thermal state a fixed point, state transformations are restricted by stronger limitations. Such processes are defined as the set of Markovian thermal processes (MTP) in Ref. \cite{PhysRevLett.129.040602,PhysRevA.106.012426}, where it is proved that any operation in MTP can be realized by a sequence of elementary thermalizations. A direct consequence of this result is that under MTP and without any auxiliary system, a two-level system cannot be cooled to a temperature lower than that of the reservoir.

By noticing that memory effect of the environment sets the difference between Markovian and non-Markovian processes, we introduce a $d$-dimensional auxiliary system, with trivial Hamiltonian and initially in a maximally mixed state, to assist state transformations under MTP. Because this auxiliary system does not provide energy or non-equilibrium resources, it acts as a ``memory'' in state transformations. 

In this paper, we study the limits on cooling and work extraction via memory-assisted Markovian thermal processes, and compare the performance with TP and MTP. For cooling a qubit system, we consider two paradigms, cooling under coherent control and cooling under incoherent control. For both paradigms, we derive the optimal ground-state populations under TP and MTP, and find that for the paradigm with coherence control, there is a gap between asymptotic ground-state populations (in the limit of infinite rounds) achieved by TP and MTP, while for the paradigm with incoherent control, both TP and MTP can reach the same asymptotic ground-state population but the convergence rate differs. We further propose memory-assisted protocols for cooling under coherent and incoherent control, which bridges the gap between the performances of TP and MTP. For the problem of extracting work from an out-of-equilibrium qubit state, we prove that if the qubit is initially in the excited state, the optimal strategy by TP can be approximated by MTP assisted by a large enough memory, indicating that memory-assisted MTP can outperform elementary thermal process when the dimension of the target is larger than 2. Our results may shed light on the study of memory effect of non-Markovian thermodynamic processes.

%Memory effect of the environment sets the different between Markovian and non-Markovian processes. pseudomodes in quantum optical processes \cite{PhysRevA.50.3650,PhysRevA.55.2290,PhysRevA.64.053813}.   srep05192,PhysRevE.84.041109,PhysRevX.7.041033

\section{Prelimitaries}
\subsection{Markovian thermal processes and related concepts}
Consider a quantum system $S$ with Hamiltonian $H_S$ and surrounded by a reservoir $R$ at inverse temperature $\beta$. In equilibrium, the state of $S$ reads $\tau\equiv e^{-\beta H_S}/\tr(e^{-\beta H_S})$, which is also called a Gibbs state.

Here we briefly review three sets of completely positive and trace-preserving (CPTP) maps: the TP \cite{PhysRevLett.115.210403,PhysRevX.5.021001},the MTP \cite{PhysRevLett.129.040602,PhysRevA.106.012426}, and ETP \cite{Lostaglio2018elementarythermal}.

A CPTP map $\operation$ belongs to TP if and only if it satisfies the following two conditions \cite{PhysRevLett.115.210403,PhysRevX.5.021001}.\\
(P1) $\operation$ is time-translationally symmetric
\begin{equation}
\operation(e^{-iH_St}\rho e^{iH_St})=e^{-iH_St}\operation(\rho) e^{iH_St},\forall \rho,t.
\end{equation}
(P2) $\operation$ preserves the Gibbs state
\begin{equation}
\operation(\tau)=\tau.
\end{equation}

Here and following, we focus on quantum states which are diagonal on the eigenbasis of Hamiltonian. Such states can be fully characterized as vectors $\boldsymbol{p}$ of occupation probabilities $p_k=\langle k|\rho|k\rangle$, where $|k\rangle$ are eigenstates of $H_S$ corresponding to energy levels $E_k$. Then the action of a CPTP map on $\rho$ can be described as a stochastic matrix acting on $\boldsymbol{p}$
\begin{equation}
\rho'=\operation(\rho)\Rightarrow \boldsymbol{p'}=G\boldsymbol p,
\end{equation}
Here $G$ is a matrix of transition probabilities $G_{k'k}=p_{k'|k}\equiv\langle k'|\operation(|k\rangle\langle k|)|k'\rangle$ from state $|k\rangle$ to state $|k'\rangle$, and $\boldsymbol{p'}$ is a vector of occupation probabilities $p'_k=\langle k|\rho'|k\rangle$ for the output state $\rho'$. 
From (P2), the population dynamics $G$ induced by TP is a stochastic matrix which preserves the Gibbs distribution. 
According to Ref. \cite{Horodecki2013thermalmajor,Korzekwa16},
\begin{equation}
    \boldsymbol{p}\stackrel{\tp}{\longrightarrow}\boldsymbol{p'}\Leftrightarrow \boldsymbol{p}\succ_{_T} \boldsymbol{p'},
\end{equation}
where $\succ_{_T}$ stands for thermo-majorization.

An elementary thermal process \cite{Lostaglio2018elementarythermal} is realized as a sequence of thermal processes which involve only two energy levels at a time. 
The set of ETP is equivalent to TP for the qubit case, while for high dimensional systems, ETP are strict subsets of TP \cite{Lostaglio2018elementarythermal}. 
When a thermal process involves only two energy levels $|i\rangle$ and $|j\rangle$ with energy gap $E_{ij}$, the reduced transition matrix in the subspace spanned by $\{|i\rangle,|j\rangle\}$ can be written as
\begin{equation}\label{eq:tp}
    G^{ij} = (1-\lambda)\iden^{ij} +\lambda\beta^{ij},
\end{equation}
where $\lambda\in[0,1]$, $\iden^{ij}$ is the identity matrix, and $\beta^{ij}$ is called $\beta$-swap between $i$ and $j$, expressed as
\begin{equation}
\beta^{ij}=
\begin{pmatrix}
  1-e^{-\beta E_{ij}}& 1\\
 e^{-\beta E_{ij}} & 0 
\end{pmatrix}.
\end{equation}
The following lemma follows directly from Eq. (\ref{eq:tp}).
\begin{lemma}\label{lemma:beta}
Consider a qubit system with Hamiltonian $H_S=E|e\rangle\langle e|$, and two states which are incoherent on energy eigenbasis with occupation vectors $\boldsymbol{p}=[p,1-p]$ and $\boldsymbol{p'}=[p',1-p']$. Then $\boldsymbol{p}$ can be transformed to $\boldsymbol{p'}$ via TP, if and only if
        \begin{equation}\label{eq:tp1}
        \left\{\begin{matrix}
            p\leq p'\le p_\beta, & p\leq\gamma,\\
           p{_\beta}\le p'\le p, & p\geq\gamma,
    \end{matrix}\right.
    \end{equation}
    where $p_\beta=[\beta^{g,e}\boldsymbol{p}]_g=1-pe^{-\beta E}$ is the ground-state population of the output state obtained from $\beta$-swap, and $\gamma\equiv1/(1+e^{-\beta E})$ is the ground-state population of the Gibbs state.
\end{lemma}

A Markovian thermal process is defined as a thermal process generated by a Markovian master equation. A Markovian process is memoryless, meaning that the evolution rate of a system at time $t$ solely depends on its state at time $t$. As recently proved in Ref. \cite{PhysRevLett.129.040602}, any state transformation induced by the set of MTP can be achieved by a sequence of elementary partial thermalizations, which involves two energy levels at a time. It follows that MTP are subsets of ETP. A partial thermalization between energy levels $E_i$ and $E_j$ ($E\equiv E_j-E_i>0$) is expressed as
\begin{equation}
T_\lambda^{ij}=
\begin{pmatrix}
  1-\lambda (1-\gamma)& \lambda\gamma\\
 \lambda (1-\gamma) & 1-\lambda\gamma
\end{pmatrix},
\end{equation}
where $\lambda \in [0,1]$.
For $\lambda=1$, $T_\lambda^{ij}$ transforms any state to a balanced distribution on $i$ and $j$. It means that if $\boldsymbol{p'}=T_1^{ij}\boldsymbol{p}$, then $p_j'=p_i'e^{-\beta E}$, $\forall\boldsymbol{p}$. This is called a full thermalization between $i$ and $j$, which we will label $T^{ij}$ in the following.

When a catalytic system is employed, the state transformations are enhanced \cite{Brandao3275CTO,PhysRevX.8.041051,PhysRevX.11.011061,PhysRevLett.128.240501,Henao2021catalytic,PhysRevA.103.022403,arXiv:2306.00798,2022arXiv220915213S,2023arXiv230313020S}.
A catalyst is finite-dimensional ancilla, which interacts with the system via free operations and then returns to the exact original state. 
When the set of free operations is TP, the initial state of a catalyst is non-equilibrium.
Interestingly, as recently discovered in Ref. \cite{2023arXiv230313020S}, with Gibbs state catalysts, elementary thermal processes can emulate any operation in TP.

% In order to study the comparisons between different sets of thermal operations, the operation cone is proposed \cite{Lostaglio2018elementarythermal}. For a given state $\rho$, the TO cone $\TOcone(\rho)$ is defined as the set of all states that can be prepared from $\rho$ under the action of TO:
% \begin{equation}
% \TOcone(\rho)=\{\rho'|\rho'=\TO(\rho)\}.
% \end{equation}
% The cones for other sets of thermal operations are defined similarly.

\subsection{Memory-assisted Markovian thermal process}
Consider a qubit system with Hamiltonian $H_S=E|e\rangle\langle e|$ and initially in state $\boldsymbol{p}=[p_0,1-p_0]$.
%The Gibbs state of this qubit at inverse temperature $\beta$ reads $\tau=\gamma|g\rangle\langle g|+(1-\gamma)|e\rangle\langle e|$ with $\gamma=1/(1+e^{-\beta E})$.
According to the results in Refs. \cite{PhysRevLett.129.040602,PhysRevA.106.012426}, $\rho$ can be transformed to state $\boldsymbol{p}^{\mathrm{MTP}}=[p^{\mathrm{MTP}},1-p^{\mathrm{MTP}}]$ by MTP, if and only if, $\gamma\le p^{\mathrm{MTP}}\leq p_0$ for $p_0>\gamma$, and  $p_0\le p^{\mathrm{MTP}}\le \gamma$ for $p_0\leq\gamma$. In other words, any state which can be reached by MTP from a qubit state $\rho$, is a mixture of $\rho$ and $\gamma$.

By employing a qubit ancillary with trivial Hamiltonian $H_A=0$ and initially in a maximally mixed state $\tau^{(2)}_A=\frac12(|1\rangle\langle 1|+|2\rangle\langle 2|)$, a global MTP can transform the state of $S$ to $\rho^{(2)}=p^{(2)}|g\rangle\langle g|+(1-p^{(2)})|e\rangle\langle e|$ with
\begin{equation}
p^{(2)}=\gamma+\gamma(1-\gamma)(\gamma-p_0).\label{eq:p2}
\end{equation}
See Appendix \ref{sec:qubit_app} for detailed calculations.
Because $p^{(2)}>\gamma$ for $p_0<\gamma$ and $p^{(2)}<\gamma$ for $p_0>\gamma$, $\rho^{(2)}$ cannot be reached from $\rho$ via MTP.
%Since the auxiliary does not provide any athermality (energy, non-equilibrium, etc.), it only serves as a ``memory'' which breaks the Markovian condition.
%This example shows that by employing a qubit memory, one can enlarge the set of states that can be reached.

It is worth mentioning that the auxiliary system considered here is similar to the catalyst system proposed in Ref. \cite{PhysRevLett.129.040602,PhysRevA.106.012426}, in the sense that the initial and final states of the auxiliary system are the same. Nevertheless, here we impose stronger requirements on the auxiliary system, i.e., trivial Hamiltonian and maximally mixed initial state. In other words, it does not provide energy or non-equilibrium resource for the state transformation of the target system. The reason why it can enable state transformations which cannot be realized by MTP is that, instead of fully thermalizing after each elementary thermalization, the assisted system memorizes its state as well as its correlation to the target system. Therefore, we will call state transformations which are assisted by such auxiliary systems the memory-assisted Markovian thermal processes.

\begin{definition}
(Memory-assisted $\mtp$, $\mmtpd$) A quantum channel $\mathcal E$ is a memory-assisted Markovian thermal process if there are an assisted system $A$ with trivial Hamiltonian $H_A=0$ and initially in a maximally mixed state $\tau_A=\iden_A/d$ and a joint Markovian thermal process $\mathcal E^{\mathrm{MTP}}$ such that
\begin{equation}
\mathcal E(\cdot)=\tr_A[\mathcal E^{\mathrm{MTP}}(\cdot\otimes\tau_A)].
\end{equation}
When the dimension of $A$ is restricted to $d$, the set of memory-assisted Markovian thermal processes is labeled as $\mmtpd$.
\end{definition}

The above definition is similar to that proposed in the recent paper \cite{arXiv:2303.12840}, but we impose the trivial Hamiltonian condition to avoid energy exchange between the memory and the other systems. The following theorem is also proved in Ref. \cite{arXiv:2303.12840}. For consistency in this paper, we provide our proof in Appendix \ref{sec:qubit_app}.

\begin{theorem}\label{th:beta_simu}
Let $\boldsymbol{p}$ be an initial state of the target system, which undergoes $\operation^{ij}\in\mmtpd$ involving only two energy levels $E_i$ and $E_j$ ($E=E_j-E_i>0$). Then the state $\boldsymbol{p^{(d)}}$ with $p_i^{(d)}$ in the following form can be reached,
    \begin{equation}\label{eq:pi_beta_simu}
        p_i^{(d)}=(1-e^{-\beta E})p_i+p_j+[(1-\gamma)p_i-\gamma p_j]\delta_d(\gamma),
    \end{equation}
    where $\delta_d(\gamma)<o[(4\gamma(1-\gamma))^d d^{-3/2}]$.
\end{theorem}
Notice that the action of $\beta$-swap $\beta^{ij}$ leads to $[\beta^{i,j}\boldsymbol{p}]_i=(1-e^{-\beta E})p_i+p_j$. Therefore, $\beta$-swap
can be simulated by $\mmtpd$ with error exponentially
decreasing in the dimension of the assisted memory.

\section{Cooling}
\subsection{Settings and main results}
Our target system is a qubit system with Hamiltonian $H_S=E|e\rangle\langle e|$ and is surrounded by a reservoir $R$ at inverse temperature $\beta$. Initially, the target system is thermalized to an equilibrium state $\tau=\gamma|g\rangle\langle g|+(1-\gamma)|e\rangle\langle e|$ with $\gamma=1/(1+e^{-\beta E})$. %This is called a Gibbs state or equilibrium state.
\begin{figure}[htbp]
\centering
\includegraphics[width=0.5\textwidth]{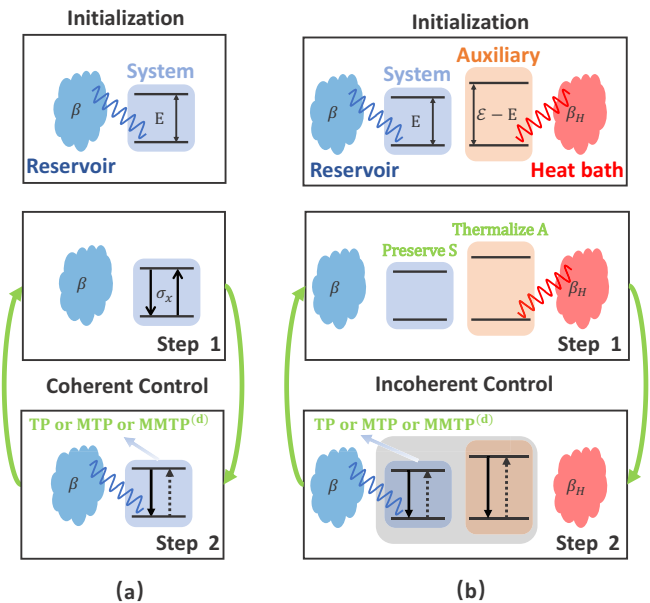}
\caption{\textbf{Cooling Paradigms.} \textbf{(a) Cooling under coherence control.} The target qubit is first initialized as a thermal state in $R$. In step 1 of each round, the target qubit transforms under the unitary $\sigma_x$; in step 2, it undergoes a process $\FO\in\fo$, where $\fo=\tp,\ \mtp,$ or $\mmtpd$. \textbf{(b) Cooling under incoherent control.} To initialize the composed system, the target qubit is thermalized in $R$ while the auxiliary is thermalized in $H$. In step 1 of each round, one preserves the state of $S$ and thermalizes $A$ in $H$; in step 2 , the composed system of $S$ and $A$ undergoes a joint process $\FO\in\fo$.}
\label{fig:cooling} 
\end{figure}

In order to cool the target system, we consider two paradigms: cooling under coherent control and cooling under incoherent control, which are schematically depicted in Fig. \ref{fig:cooling}.
Both paradigms consist of $n$ rounds. In each round, the target system absorbs the energy for cooling and then undergoes a process which belongs to TP, MTP, or $\mmtpd$. The difference between the two paradigms lies in the source of energy. 
In the paradigm of coherent control, the energy for cooling comes from a unitary operation, while the energy in incoherent control is provided by a hot bath $H$ at inverse temperature $\beta_H<\beta$.

In the paradigm of coherent control, each round consists of two steps. In the first step,  a unitary operation is applied to invert the populations of the target system, while the target system is decoupled from the heat bath. The second step is to maximize the ground-state population via TP, MTP or $\mmtpd$. We denote the state of the target qubit after the $n$th cycle as $\rho_{n}^{\fo,coh}=p_{n}^{\mathrm{F},coh}|g\rangle\langle g|+(1-p_{n}^{\mathrm{F},coh})|e\rangle\langle e|$, where $\mathrm{F}=\tp,\ \mtp$, or $\mmtpd$ denotes the allowed operations in the cooling process, and the superscript $coh$ means that the energy comes from coherent control.
% we denote the state of the target system before the $n$th cycle as $\rho_{n-1}^{\fo,coh}=p_{n-1}^{\mathrm{F},coh}|g\rangle\langle g|+(1-p_{n-1}^{\mathrm{F},coh})|e\rangle\langle e|$, where $\mathrm{F}=\tp,\ \mtp$, or $\mmtpd$ denotes the allowed operations in the cooling process, and the superscript $coh$ means that the energy comes from coherent control.
Our main result for coherent control is summarized in the following theorem.
\begin{theorem}\label{th:1}
    Under coherent control, the ground-state populations after the $n$th round, under  cooling processes involving $\tp$ and $\mtp$, are respectively upper bounded by
    \begin{eqnarray}
        p_{n,*}^{\tp,coh}&=&1-(1-\gamma)e^{-n\beta E},\label{eq:pntp}\\
        p_{n,*}^{\mtp,coh}&=&\gamma,
    \end{eqnarray}
    where the bounds are reachable.
    Moreover, under $\mmtpd$, the ground-state population can reach
    \begin{equation}
        p_{n,*}^{(d),coh}=p_{\max}^{(d),coh}-(e^{-\beta E}-\delta_d(\gamma))^n(p_{\max}^{(d),coh}-\gamma),
    \end{equation}
    where
    \begin{equation}\label{eq:pdmax}
        p_{\max}^{(d),coh}=1-\frac{\gamma }{1+(1-e^{-\beta E})/\delta_d(\gamma)}.
    \end{equation}
\end{theorem}
Detailed proof and discussion of the above theorem are in Sec. \ref{subsec:uni_con}.

In the paradigm of incoherent control, we employ an auxiliary qubit with Hamiltonian $H_A=(\mE-E) |1_A\rangle\langle 1_A|$, where we set $\mE>E$. 
Initially, the target system $S$ is thermalized in $R$ while the auxiliary is thermalized in $H$. Each round of the protocol consists of two steps.
In the first step, the auxiliary is fully thermalized in $H$, absorbing the energy for cooling, while the state of target system is preserved. 
In the second step, the target system and the auxiliary are brought in contact with $R$ and undergo a thermal process.
After the $n$th round, the state of the target system becomes
$\rho_{n}^{\fo,inc}=p_{n}^{\mathrm{F},inc}|g\rangle\langle g|+(1-p_{n}^{\mathrm{F},inc})|e\rangle\langle e|$, where the superscript $inc$ indicates incoherent control.
We prove the following.
\begin{theorem}\label{th:2}
After the $n$th round, ground-state population of the target system under incoherent control can reach
\begin{equation}\label{eq:pnfinc}
p_n^{\fo,inc}=p_{*}^{inc}-v_\fo^n(p_{*}^{inc}-\gamma),
\end{equation}
$\forall \fo=\tp,\mtp,\mmtpd$. the asymptotic ground-state population reads
\begin{equation}
p_{*}^{inc}=\frac{1}{1+e^{-\beta\mE}e^{\beta_H(\mE-E)}}.\label{eq:pmax}
\end{equation}
The convergence rates for $\fo=\tp,\mtp$, and $\mmtpd$ are respectively
\begin{eqnarray}
   v_\tp&=&\eta(1-e^{-\beta\mE}),\label{eq:vtp_inc}\\
   v_\mtp&=&v_\tp+\frac{1-\eta+\eta e^{-\beta\mE}}{1+e^{\beta\mE}},\label{eq:vmtp_inc}\\
   v_{\mmtpd}&=&v_\tp+\frac{\eta-(2\eta-1)}{1+e^{-\beta\mE}}\delta_d(\gamma_\mE),\label{eq:vmmtp_inc}
\end{eqnarray}
where $\gamma=1/(1+e^{-\beta E})$, $\eta=1/(1+e^{-\beta_H (\mE-E)})$, and $\gamma_\mE=1/(1+e^{-\beta\mE})$.
Further, for $\fo=\tp,\mtp$, Eq. (\ref{eq:pnfinc}) is also the upper bound for ground-state population.
\end{theorem}
It can be inferred from the theorem that, cooling processes under $\tp$, $\mtp$, and $\mmtpd$ can all approach $p_{*}^{inc}$ in the asymptotic limit, while the rates of convergence are different. Further, for $H$ in the high temperature limit $\beta_H\rightarrow0$, the asymptotic ground-state population $p_{*}^{inc}$ reaches its maximum. Then we arrive at the following corollary.
\begin{corollary}
Under incoherent control, in the limit of infinity cycles, the ground-state population of a qubit target system is upper bounded by
\begin{equation}
    p_*=\frac{1}{1+e^{-\beta\mE}}.
\end{equation}
where $\mE$ stands for the largest energy gap of the composed system consisting of the target qubit $S$ and the auxiliary qubit $A$.
\end{corollary}

\subsection{Cooling under coherence control}\label{subsec:uni_con}
 In the $n$th round,  the state of the system first transforms under the unitary $\sigma_x:=|g\rangle\langle e|+|e\rangle\langle g|$, and then undergoes a process $\FO\in\fo$. Precisely, the ground-state population after the $n$th round can be expressed as
\begin{eqnarray}
p_n^{\fo,coh}&=&\langle g|\FO\left((1-p_{n-1}^{\fo,coh})|g\rangle\langle g|+p_{n-1}^{\fo,coh}|e\rangle\langle e|\right)|g\rangle,\nonumber\\
&=&p^{\fo}_{g|g}+(p^{\fo}_{g|e}-p^{\fo}_{g|g})p_{n-1}^{\fo,coh}.
\end{eqnarray}
%where $p^{\fo}_{g|g}:=\langle g|\FO(|g\rangle\langle g|)|g\rangle$ is the transition probability, and $p^{\fo}_{g|e}$ is defined similarly.
Because $p_{0}^{\fo,coh}=\gamma$,
it follows that
\begin{equation}\label{eq:pnf}
    p_n^{\fo,coh}=p_*^{\fo,coh}-v^n_{\fo,coh}(p_*^{\fo,coh}-\gamma),
\end{equation}
where $p_*^{\fo,coh}=p^{\fo}_{g|g}/(p^{\fo}_{g|g}+p^{\fo}_{e|e})$ and $v_{\fo,coh}=p^{\fo}_{g|e}-p^{\fo}_{g|g}$.
Because $|v_{\fo,coh}|<1$ as long as $\FO$ is not an identical operation, the ground-state population converges to $p_*^{\fo,coh}$ at a speed exponentially fast in the number of rounds.

For $\fo=\tp$, Eq. (\ref{eq:pntp}) has been proved in \cite{Alhambra2019heatbathalgorithmic}. For consistency, we give the proof as follows. From Eq. (\ref{eq:tp}), we have
\begin{equation}
    p_*^{\tp,coh}=(1+\frac{1-\lambda}{1-\lambda e^{-\beta E}})^{-1},
\end{equation}
and $v_{\tp,coh}=\lambda(1+e^{-\beta E})-1$ with $\lambda\in[0,1]$.
It follows that if one requires to cool the system to a temperature below that of the reservoir, i.e., $p_*^{\tp,coh}\geq \gamma$, the thermal process employed should satisfy $\lambda\geq\gamma$.
Further, from Lemma \ref{lemma:beta}, in each round, $p_n^{\tp,coh}$ is maximized at $\lambda=1$, as long as $1-p_{n-1}^{\tp,coh}\leq \gamma$, which naturally holds. For $\lambda=1$, we have $p_*^{\tp,coh}=1$ and $v_{\tp,coh}=e^{-\beta E}$. It then follows from Eq. (\ref{eq:pnf}) that the maximum of $p_n^{\tp,coh}$ is in the form of Eq. (\ref{eq:pntp}). A direct consequence is that, under coherence control and with $\fo=\tp$, the target qubit can be cooled to absolute zero in the asymptotic limit .

In contrast, when $\fo=\mtp$, the system can only be cooled to the temperature of the reservoir. This is because MTP acting on a qubit system is equivalent to partial thermalizations.

When $\fo=\mmtpd$, we apply simulated $\beta$-swap on the target qubit and thermalize the memory before each round.
The ground-state population after the $n$th round can be calculated by using Eq. (\ref{eq:pi_beta_simu}) and setting $p_i=1-p_{n-1}^{(d),coh}$ and $p_j=1-p_i$. Precisely, We obtain the following
\begin{eqnarray}
p^{(d),coh}_n&=&[e^{-\beta E}-\delta_d(\gamma)]p_{n-1}^{(d),coh}\nonumber\\
&&+1-e^{-\beta E}+(1-\gamma)\delta_d(\gamma)\nonumber\\
&=& p_{\max}^{(d),coh}-[e^{-\beta E}-\delta_d(\gamma)](p_{\max}^{(d),coh}-p_{n-1}^{(d),coh})\nonumber\\
&=& p_{\max}^{(d),coh}-[e^{-\beta E}-\delta_d(\gamma)]^n(p_{\max}^{(d),coh}-\gamma),
\end{eqnarray}
where $p_{\max}^{(d),coh}$ is in the form of Eq. (\ref{eq:pdmax}).
This completes the proof of Theorem \ref{th:1}.

Careful analysis shows that $p_{\max}^{(d),coh}>\gamma$, $\forall d\geq2$, indicating that cooling is enhanced by the memory. Further, $p_{\max}^{(d),coh}$ is monotonically increasing in $d$, which means that the enhancement grows with the size of the memory. Last but not least, $p_{\max}^{(1),coh}=p_{*}^{\mtp,coh}=\gamma$ and $p_{\max}^{(\infty),coh}=p_{*}^{\tp,coh}=1$, and this bridges the gap between the performances of TP and MTP.

\subsection{Cooling under incoherent control}
Initially, the target qubit is fully thermalized in $R$ while the auxiliary is fully thermalized in $H$.
On basis $\{|g0_A\rangle,|g1_A\rangle,|e0_A\rangle,|e1_A\rangle\}$,
the initial state of $SA$ can be written as
\begin{eqnarray}
\boldsymbol{p^{(0)}}&=&[p^{(0)}_{g0},p^{(0)}_{g1};p^{(0)}_{e0},p^{(0)}_{e1}]\nonumber\\
    &=&[\gamma\eta,\gamma(1-\eta);(1-\gamma)\eta,(1-\gamma)(1-\eta)],
    %\nonumber\\
%    &=&\gamma\eta[1,e^{-\beta_H(\mE-E)},e^{-\beta E},e^{-\beta_H\mE}e^{-(\beta-\beta_H)E}],
\end{eqnarray}
where $\eta=1/(1+e^{-\beta_H(\mE-E)})$.
The target in the first round is then to maximize the sum of the occupations on $|g0_A\rangle$ and $|g1_A\rangle$ by a quantum operation in $\fo$.

We first consider the case $\fo=\tp$. 
Let $\boldsymbol{\tau}=[\gamma,1-\gamma]\otimes[\gamma_A,1-\gamma_A]$ with $\gamma_A=1/(1+e^{-\beta(\mE-E)})$ be the Gibbs state of $SA$ in the reservoir $R$.
Because $\beta_H<\beta$, we have
\begin{equation}
    \frac{p^{(0)}_{g1}}{\tau_{g1}}=\frac{p^{(0)}_{e1}}{\tau_{e1}}>\frac{p^{(0)}_{g0}}{\tau_{g0}}=\frac{p^{(0)}_{e0}}{\tau_{e0}}.
\end{equation}
It means that one can neither increase the population on $|g1_A\rangle$, nor decrease the population on $|e0_A\rangle$ via any operation in TP. The optimal strategy is then the $\beta$-swap between the $|g0_A\rangle$ and $|e1_A\rangle$.
The ground-state population of $S$ then becomes
\begin{equation}
    p_1^{\tp,inc}=1-\eta+\gamma\eta(1-e^{-\beta\mE}).
\end{equation}
Equivalently,
\begin{equation}\label{eq:p1inc_recurrence}
    (p_{*}^{inc}-p_1^{\tp,inc})=v_{\tp}(p_{*}^{inc}-\gamma),
\end{equation}
where $p_{*}^{inc}$ and $v_{\tp}$ are defined in Eqs. (\ref{eq:pmax}) and (\ref{eq:vtp_inc}) respectively. Notice that $p_{*}^{inc}>p_1^{\tp,inc}$, because $p_{*}^{inc}>\gamma$ and $v_{\tp}>0$.

In the second round, the auxiliary is first fully thermalized in $H$, resulting in state $[\eta,1-\eta]$. The joint state of $SA$ becomes
\begin{eqnarray}
\boldsymbol{p^{(1)}}&=&[p^{(1)}_{g0},p^{(1)}_{g1};p^{(1)}_{e0},p^{(1)}_{e1}]\nonumber\\
    &=&[p_1^{\tp,inc}\eta,p_1^{\tp,inc}(1-\eta);\nonumber\\
    &&(1-p_1^{\tp,inc})\eta,(1-p_1^{\tp,inc})(1-\eta)].
\end{eqnarray}
Then we check that
\begin{equation}\label{eq:order}
    \frac{p^{(1)}_{g1}}{\tau_{g1}}>\frac{p^{(1)}_{e1}}{\tau_{e1}}>\frac{p^{(1)}_{g0}}{\tau_{g0}}>\frac{p^{(1)}_{e0}}{\tau_{e0}}.
\end{equation}
Therefore, the best strategy is still the $\beta$-swap between the $|g0_A\rangle$ and $|e1_A\rangle$. The resulted ground-state population $p_2^{\tp,inc}$ then satisfies
\begin{equation}
    (p_{*}^{inc}-p_2^{\tp,inc})=v_{\tp}(p_{*}^{inc}-p_1^{\tp,inc}).
\end{equation}
We apply the above discussion $n$ times and obtain the recurrence relation
\begin{equation}\label{eq:pninc_recurrence}
    (p_{*}^{inc}-p_n^{\tp,inc})=v_{\tp}(p_{*}^{inc}-p_{n-1}^{\tp,inc}).
\end{equation}
Importantly, this recurrence relation implies $p_{*}^{inc}>p_n^{\tp,inc}$ as long as $p_{*}^{inc}>p_{n-1}^{\tp,inc}$, which in turn ensures that the ordering in Eq. (\ref{eq:order}) holds for $p^{(n)}$. This is the reason for employing the $\beta$-swap between $|g0_A\rangle$ and $|e1_A\rangle$ as the optimal cooling strategy in every round.
Combining Eqs. (\ref{eq:p1inc_recurrence}) and (\ref{eq:pninc_recurrence}), we arrive at Eq. (\ref{eq:pnfinc}) in Theorem \ref{th:2} for $\fo=\tp$.

As for $\fo=\mtp$, the discussions are similar. The only difference is that $\beta$-swap is not in $\mtp$. The optimal strategy in each cycle then becomes full thermalization between  $|g0_A\rangle$ and $|e1_A\rangle$. Therefore, we have the following recurrence relation
\begin{eqnarray}
    p_n^{\mtp,inc}&=&\gamma[p_{n-1}^{\mtp,inc}\eta+(1-p_{n-1}^{\mtp,inc})(1-\eta)]\nonumber\\
    &&+(1-\eta)p_{n-1}^{\mtp,inc}.
\end{eqnarray}
where $p_0^{\mtp,inc}=\gamma$.
This is equivalent to
\begin{equation}
    (p_{*}^{inc}-p_n^{\mtp,inc})=v_{\mtp}(p_{*}^{inc}-p_{n-1}^{\mtp,inc}).
\end{equation}
Equation (\ref{eq:pnfinc}) then follows directly for $\fo=\mtp$.
Interestingly, even though the allowed thermal process is restricted to be Markovian, one can still approach $p_{*}^{inc}$ in the asymptotic limit. The difference between the performances of $\tp$ and $\mtp$ is the convergence rate. Because $v_{\mtp}>v_{\tp}$, $p_n^{\tp,inc}$ converges faster than $p_n^{\mtp,inc}$.

For $\fo=\mmtpd$, our strategy is to implement simulated $\beta$-swap between $|g0_A\rangle$ and $|e1_A\rangle$ in each cycle.
From Theorem \ref{th:beta_simu}, where we set $p_i=\eta p_{n-1}^{(d),inc}$ and $p_j=(1-\eta)(1-p_{n-1}^{(d),inc})$, the sum of populations on $|g0_A\rangle$ and $|g1_A\rangle$ after the $n$th round then reads
\begin{eqnarray}
    p_{n}^{(d),inc}&=&(1-\eta)p_{n-1}^{(d),inc}+(1-e^{-\beta\mE})\eta p_{n-1}^{(d),inc}\nonumber\\
    &&+(1-\eta)(1-p_{n-1}^{(d),inc})[(1-\gamma_{\mE})\eta p_{n-1}^{(d),inc}\nonumber\\
    &&-\gamma_{\mE}(1-\eta)(1-p_{n-1}^{(d),inc})]\delta_d(\gamma_{\mE}),\nonumber\\
    &=&p_{*}^{inc}-v_{\mmtpd}(p_{*}^{inc}-p_{n-1}^{(d),inc}),\nonumber\\
    &=&p_{*}^{inc}-v_{\mmtpd}^n(p_{*}^{inc}-\gamma),
\end{eqnarray}
with $v_{\mmtpd}$ in the form of Eq. (\ref{eq:vmmtp_inc}). This completes the proof for $\fo=\mmtpd$.
Still, the asymptotic limit of $p_{n}^{(d),inc}$ is the same as that of $p_n^{\tp,inc}$ and $p_n^{\mtp,inc}$. Further, $v_{\mmtpd}$ is monotonically decreasing in $d$. It bridges the gap between the convergence rates for $\tp$ and $\mtp$, in the sense that $v_{\mathrm{MMTP}^{(1)}}=v_{\mtp}$ and $v_{\mathrm{MMTP}^{(\infty)}}=v_{\tp}$.

\section{Work extraction}
Consider a qubit $S$ initially in a non-equilibrium state $[p_0,1-p_0]$, and a work bit with Hamiltonian $H_{_\W}=\W|1\rangle\langle 1|$ initially in the ground state $[1,0]$.
The task of work extraction is to transform the state of the work bit to its exited state $|1\rangle$ by joint thermal processes on $SW$.
Precisely, in a single-shot work extraction,
\begin{equation}
[p_0,1-p_0]_S\otimes[1,0]_{\W} \stackrel{\fo}{\longmapsto} [\gamma,1-\gamma]_S\otimes [\epsilon,1-\epsilon]_{\W},
\end{equation}
where $\fo=\tp,\mtp,\etp$, or $\mmtpd$, and $\epsilon$ is called the error of work extraction. The task of work extraction is to minimize $\epsilon$ over $\fo$.

Here we analytically solve the above problem for the case where the target qubit is in the excited state, i.e., $p_0=0$, and prove the following theorem.
\begin{theorem}\label{th:3}
    Consider a qubit $S$ with Hamiltonian $H_S=E|e\rangle\langle e|$ and initially in state $[0,1]_S$, and a work bit with Hamiltonian $H_{_W}=W|1\rangle\langle 1|$ and initially in state $[1,0]_W$. By $\mmtpd$, the error $\epsilon^{(d)}$ of work extraction can reach
    \begin{equation}
        \epsilon^{(d)}(W)=I_d(\frac{1}{1+e^{-\beta(E-W)}},\frac{1}{1+e^{\beta W}}).
    \end{equation}
    Moreover, for $d\rightarrow\infty$, the error can approach
    \begin{eqnarray}
        &&\epsilon^{(\infty)}(W)=\epsilon^{\tp}(W)\nonumber\\
        &=&
\left\{\begin{matrix}
  0, & \W \le W_0, \\
  1-e^{\beta(E-W)}-e^{-\beta\W},   & \W > W_0. 
\end{matrix}\right.\label{eq:epw}
    \end{eqnarray}
    where $Z\equiv 1+e^{-\beta E}$ is the partition function of the target qubit, $W_0\equiv E+k_BT\ln Z$ is the maximal work which can be extracted from $S$ via $\tp$ with vanishing error, and $\epsilon^{\tp}(W)$ is the minimum error that can be reached via $\tp$ for given $W$.
\end{theorem}
Here the function $I_d(x,y)$ is defined in Eq. (\ref{eq:Id}) in Appendix \ref{sec:app1}.
% \begin{equation}
% I_d(x,y)=\frac{(1-x)^d(1-y)^d}{d}\sum^{d-1}_{j=0}\sum^{d-j-1}_{k=0}(d-j-k)f^{(k)}_{d}x^{k}f^{(j)}_{d}y^j,
% \end{equation}
% where $f_d^{(k)}\equiv C_{d+k-1}^k$ and $C_m^n:=\frac{n!}{m!(n-m)!}$.

This theorem provides an evidence in the regime of high-dimensional systems for the conjecture that $\mmtpd$ can approach TP for $d\rightarrow\infty$. This conjecture is not trivial, because if it is proved, then the thermal bath can be divided into two parts. The first one can exchange energy with the system but does not have memory effect, while the second one serves as a memory and does not exchange energy with the system.

\subsection{Error of work extraction via TP, MTP and ETP}
In the following, we derive the minimum error of work extraction as a function of $W$, when the allowed operations are TP, MTP and ETP. In particular, for comparison with the performance of $\mmtpd$, we focus on the case with $p_0=0$. On basis $\{|g0\rangle,|g1\rangle,|e0\rangle,|e1\rangle\}$, the initial state of $SW$ then reads $\boldsymbol{p}_0=[0,0,1,0]$. The task of work extraction is then to minimize the sum of populations on $|g0\rangle$ and $|e0\rangle$ in the output by the allowed thermal processes.

First, we consider the case with $\fo=\tp$. The optimal extraction error $\epsilon^{\tp}(W)$ can be computed via the thermo-majorization condition
\begin{equation}
    [0,1]_S\otimes[1,0]_{\W} \succ_{_T} [\gamma,1-\gamma]_S\otimes [\epsilon,1-\epsilon]_{\W}.
\end{equation}

This condition leads to
\begin{equation}
 \epsilon_{\tp}(W)=
\left\{\begin{matrix}
  0, & \W \le W_0, \\
  1-e^{\beta(E-W)}-e^{-\beta\W},   & \W > W_0.
\end{matrix}\right.
\end{equation} 
where $W_0$ is defined in theorem \ref{th:3}. The optimal strategy is to implement a joint thermal operation described by the transition matrix $G^\tp$, followed by a local full thermalization on $S$. The form of $G^\tp$ reads as follows.\\
For $W\leq E$,
\begin{equation}
G^\tp_1=\left(
    \begin{matrix}
        1 & 0 & 0  & 0\\
        0 & 1-e^{-\beta(E-W)} & 1 & 0\\
        0 & e^{-\beta(E-W)} & 0 & 0\\
        0 & 0 & 0 & 1
    \end{matrix}\right).
\end{equation}
For $E< W\leq W_0$, which implies $e^{\beta E}< e^{\beta W} \leq 1+e^{\beta E}$,
\begin{equation}
G^\tp_2=\left(
    \begin{matrix}
        1 & 0 & 0  & 0\\
        0 & 0 & e^{-\beta(W-E)} & 0\\
        0 & 1 & 0 & e^{\beta W}-e^{\beta E}\\
        0 & 0 & 1-e^{-\beta(W-E)} & 1-e^{\beta W}+e^{\beta E}
    \end{matrix}\right).\label{eq:g2tp}
\end{equation}
For $W> W_0$, which implies $ e^{\beta W} > 1+e^{\beta E}$,
\begin{equation}
G^\tp_3=\left(
    \begin{matrix}
        1 & 0 & 0  & 0\\
        0 & 0 & e^{-\beta(W-E)} & 0\\
        0 & 1 & 1-e^{-\beta W}(1+e^{\beta E}) & 1\\
        0 & 0 & e^{-\beta W} & 0
    \end{matrix}\right).\label{eq:g3tp}
\end{equation}
The above form of $G^\tp$ minimizes the sum of $G_{g0|e0}$ and $G_{e0|e0}$, and thus minimizes $\epsilon$ in the output.

Next we study the performance of MTP. The optimal strategy reads
\begin{equation}
    G^\mtp=T_ST^{g1,e0}T^{g1,e1},
\end{equation}
or
\begin{equation}
    \tilde{G}^\mtp=T_ST^{g1,e1}T^{g1,e0},
\end{equation}
where $T_S$ denotes full thermalization of system $S$.
Both $G^\mtp$ and $\tilde{G}^\mtp$ give
\begin{equation}
    \epsilon_{\mtp}(W)=\frac{1}{(1+e^{\beta(E-\W})(1+e^{-\beta\W})}.
\end{equation}
Notice that, for $\W \to 0$, $\epsilon_{\mtp}(W)  \to \frac{1}{2}(1-\gamma) \ne 0$.

For $\fo=\etp$, we find that the optimal transition matrix is
\begin{equation}
    G^\etp=T_S\beta^{g1,e0}\beta^{g1,e1},
\end{equation}
or
\begin{equation}
    \tilde{G}^\etp=T_S\beta^{g1,e1}\beta^{g1,e0}.
\end{equation}
It then follows that
\begin{equation}
 \epsilon_{\etp}(W)=
\left\{\begin{matrix}
  0, &  \W \le E,\\
  (1-e^{\beta\delta})(1-e^{-\beta\W}),   &  \W > E.
\end{matrix}\right.
\end{equation}

% the optimal error occurs under $\beta$-swap, if $E \ge \W $, $\epsilon_{ETP}=0$. If $E< \W $
% \begin{equation}
% \boldsymbol{q}=\beta^{E,E+W}\beta^{E,W}\boldsymbol{p}\\
% \end{equation}
% So
% \begin{eqnarray}
%  \nonumber \epsilon_{ETP} &=& (1-e^{\beta\delta})(1-e^{-\beta\W})\\
%                  &=& \epsilon_{\tp}+e^{\beta(E-2\W)},
% \end{eqnarray}

%Fig.(\ref{fig:error_W}) shows that $\mmtpt$ obviously reduces the error compare with MTP and $\mmtptt$ almost overlaps with $\tp$ except the region around the inflection point $W0$.

\begin{figure}[htbp] 
\centering 
\includegraphics[width=0.50\textwidth]{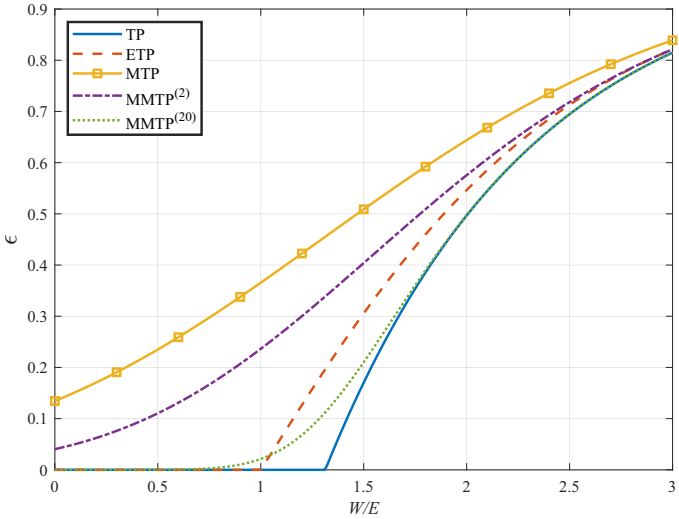} 
\caption{ \textbf{The error \texorpdfstring {$\bm{\epsilon}$}{} of work extraction as a function of extracted work \texorpdfstring {$\bm{W}$}{}.} Work is extracted from  a qubit $S$ with energy gap $E$ and initially in the  excited state $[0,1]_S$.} 
\label{fig:error_W} 
\end{figure}

Here We briefly mention that the Gibbs-preserving stochastic matrices $G_2^\tp$ and $G_3^\tp$ in Eqs. (\ref{eq:g2tp}) and (\ref{eq:g3tp}) satisfy the detailed balanced condition, i.e., $G_{j|i}=e^{-\beta(E_j-E_i)}G_{i|j}$. However, these transformations are impossible via ETP, because as shown in Fig. \ref{fig:error_W}, the error of work extraction achieved by these two stochastic matrices cannot be reached by ETP. This observation shows that detailed balance is a necessary but not sufficient condition for a thermal process to be realized as a sequence of thermal processes which involve only two energy levels at a time.

\subsection{Memory-assisted protocol for work extraction}\label{sec:protocol}
\begin{figure}[htbp] 
\centering
\includegraphics[width=0.5\textwidth]{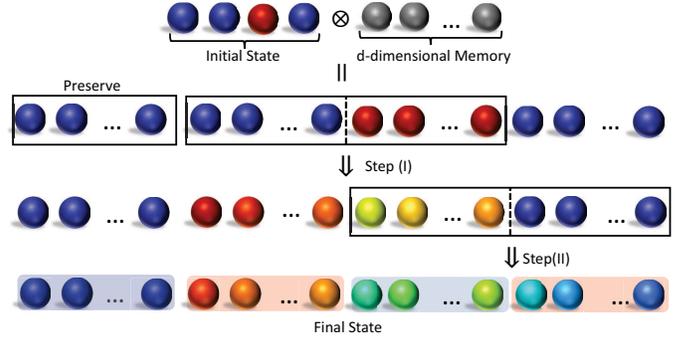} 
\caption{\textbf{Scheme for work extraction under \texorpdfstring {$\bm{\mmtpd}$}{}.} 
The  color of each dot stands for the population distribution.
Initially, the qubit $S$ is in the excited state and thus the composed state of $SW$ reads $[0,0,1,0]_{SW}$, while the $d$-dimensional memory is in a maximally mixed state.
In step (I), MTP is applied to the subspace spanned by $\{|g11\rangle,\dots,|g1d\rangle;|e01\rangle,\dots,|e0d\rangle\}$.
In step (II), MTP is applied to the subspace spanned by $\{|e01\rangle,\dots,|e0d\rangle;|e11\rangle,\dots,|e1d\rangle\}$.
The error $\epsilon^{(d)}$ of extraction equals to the sum of populations on $|e01\rangle,\dots,|e0d\rangle$.} 
\label{fig:nmemory} 
\end{figure}

Here we propose a protocol for work extraction from a qubit system in the excited state with the assistance of a $d$-dimensional memory.
The basis of the Hilbert space of the composed system $SWM$ reads $\{|g\rangle,|e\rangle\}_S\otimes\{|0\rangle,|1\rangle\}_W\otimes\{|1\rangle,\dots,|d\rangle\}_M=\{|\xi\zeta k\rangle\}_{\xi=g,e;\zeta=0,1;k=1,\dots,d}$.
Initially, the system is in the excited state $|e\rangle$, the work bit is in the ground state $|0\rangle$, and the memory is in a maximally mixed state $\frac{\iden}{d}$.
Our protocol is schematically depicted in Fig. \ref{fig:nmemory}.

Precisely, our protocol consist of two steps, which are summarized as
\begin{eqnarray}
  \nonumber &&[0,0,1,0]\otimes[\overbrace{\frac{1}{d},...,\frac{1}{d}}^{d \ times}] \\
 % \nonumber && =[\overbrace{0,...,0}^d,\overbrace{0,...,0}^d,\overbrace{\frac{1}{d},...,\frac{1}{d}}^d,\overbrace{0,...,0}^d]\\
   \nonumber && =[{0,...,0};0,...,0;\frac{1}{d},...,\frac{1}{d};0,...,0]\\
  \nonumber && \stackrel{\textbf{(\Rmnum{1})}}{\longmapsto} [0,...,0;a_d^{(1)},...,a_d^{(d)};b_1^{(d)},...,b_d^{(d)};0,...,0]\\
\nonumber  &&\stackrel{\textbf{(\Rmnum{2})}}{\longmapsto} [0,...,0;a_d^{(1)},...,a_d^{(d)};\epsilon_1,...,\epsilon_d;c_1^{(d)},...,c_d^{(d)}].
\end{eqnarray}
In step (I), simulated $\beta$-swap is implemented between $|e\rangle_S\otimes |0\rangle_W$ and $|g\rangle_S\otimes |1\rangle_W$. Notice that instead of fully thermalizing $M$ after the above operation, we preserve the total state of $SWM$ for the second step.
By combining Eqs. (\ref{eq:bjk}), (\ref{eq:ajk}) and (\ref{eq:sjk}) together in Appendix \ref{sec:nd}, we obtain
\begin{eqnarray}
b_j^{(d)}&=&\frac{(1-\gamma_\delta)^d}{d}\sum^{j-1}_{j'=0}f^{(j')}_{d}\gamma_\delta^{j'},\label{eq:bdj}
\end{eqnarray}
where $\delta=W-E$, and $\gamma_\delta=1/(1+e^{-\beta\delta})$. See Appendix \ref{sec:proof} for detailed derivation. Notice that $b^{(d)}_j$ is monotonically increasing with $j$.

Step (II) consists of $d$ subroutines. In the $k$th subroutine, full elementary thermalizations are sequentially implemented between $|e0k\rangle$ and $|e1j\rangle$ with $j=1,\dots,d$. This results in
\begin{equation}
    \epsilon_k=\sum_{j=1}^k b_{k+j-1}^{(d)} \gamma_W^d(1-\gamma_W)^{j-1}f_d^{(j-1)}.
\end{equation}
Consequently, we have
\begin{eqnarray}
\epsilon^{(d)}=\sum^d_{k=1}\epsilon_k =I_d(\gamma_\delta,1-\gamma_W).
\end{eqnarray}
The detailed derivation of the above formulas can be found in Appendix \ref{sec:proof}. Further, in Appendix \ref{sec:app1}, we prove that
\begin{eqnarray}
 &&\lim_{d \to \infty}  I_d(x,y)\nonumber\\
 &=&
\left\{\begin{matrix}
  1-\frac{x}{1-x}-\frac{y}{1-y}, & \frac{x}{1-x} +\frac{y}{1-y} < 1,\\
 0,   & \frac{x}{1-x} +\frac{y}{1-y} \ge 1.
\end{matrix}\right.
\end{eqnarray} 
This directly leads to Eq. (\ref{eq:epw}) in Theorem \ref{th:3}.
The comparison between the performances of TP, MTP, ETP and $\mmtpd$ with finite $d$ is plotted in Fig. \ref{fig:error_W}. Interestingly, although $\epsilon^{(d)}(W)$ with finite $d$ is an analytic function, in the limit of $d\rightarrow\infty$, it approaches $\epsilon^{\tp}(W)$, whose first derivative on $W$ is not continuous at $W=W_0$.

\section{conclusion}
We have introduced a finite-dimensional memory system with trivial Hamiltonian and initially in a maximally mixed state to bridge the gap between the performances of Markovian thermal processes and general thermal processes in thermodynamic tasks, including cooling and work extraction. We study two cooling paradigms, under coherent control and under incoherent control. In both paradigms, the ground-state population of the target qubit approaches a limit exponentially fast  in the number of rounds, whether the set of allowed processes is TP, MTP or $\mmtpd$. For the paradigm under coherence control, the limit of achievable ground-state population via $\mmtpd$ is monotonically increasing in $d$, which reduces to the limit for MTP when $d=1$, and approaches the limit for TP when $d\rightarrow\infty$. For the paradigm under incoherent control, we show that the limits of achievable ground-state population are the same for TP, MTP, and $\mmtpd$, while the convergence rates are different. Still, the rate for $\mmtpd$ bridges the rates for TP and MTP.

In the problem of work extraction, the system which undergoes $\mmtpd$ consists of an out-of-equilibrium qubit $S$ and a work qubit $W$. We derive the analytic expression for the error of work extraction achieved by $\mmtpd$ when $S$ is initially in the excited state. Interestingly, the extraction error achieved by $\mmtpd$, which is an analytic function of the energy gap $W$ of the work bit, approaches in the limit $d\rightarrow\infty$ to the extraction error achieved by TP, whose first-order derivation is not continuous.

Our results provide evidences for the conjecture that $\mmtpd$ can simulate transformations under TP with vanishing error, when the size $d$ of the memory is large enough.

\begin{acknowledgments}
This work was supported by NSFC under Grant No. 11774205.
\end{acknowledgments}

\newpage
\begin{widetext}
% \begin{equation}
% {\cal R}^{(\text{d})}=
%  g_{\sigma_2}^e
%  \left(
%    \frac{[\Gamma^Z(3,21)]_{\sigma_1}}{Q_{12}^2-M_W^2}
%   +\frac{[\Gamma^Z(13,2)]_{\sigma_1}}{Q_{13}^2-M_W^2}
%  \right)
%  + x_WQ_e
%  \left(
%    \frac{[\Gamma^\gamma(3,21)]_{\sigma_1}}{Q_{12}^2-M_W^2}
%   +\frac{[\Gamma^\gamma(13,2)]_{\sigma_1}}{Q_{13}^2-M_W^2}
%  \right)\;. \label{eq:wideeq}
% \end{equation}
\appendix
% \section*{Appendices}
\section{SEVERALe FUNCTIONS}\label{sec:app1}
In this appendix, we give the definitions to several functions, and derive some equivalent expressions for these functions.
\begin{definition}
%\item
We define functions $L_n^{(m)}(x)$, $K_n^{(m)}(x)$, and $I_n^{(m)}(x)$ as follows:
\begin{eqnarray}
L_n^{(m)}(x) & := & (1-x)^n\sum_{j=0}^m f_n^{(j)}x^j,\\
K_n^{(m)}(x) & := & \frac{(1-x)^n}{n}\sum_{j=0}^m jf_n^{(j)}x^j,\\
I_n^{(m)}(x) & := & \frac{(1-x)^n}{n}\sum_{j=0}^m (n-j)f_n^{(j)}x^j,\label{eq:Ix}
\end{eqnarray}
where $x\in[0,1]$, $m$ and $n$ are positive integers satisfying $m\leq n-1$, and $f_n^{(j)}$ are defined by the following recurrence relation
\begin{equation}
    f_j^{(0)}=1,\ f_j^{(k+1)}=\sum_{j'=1}^j f_{j'}^{(k)},
\end{equation}
or equivalently, 
\begin{equation}
    f_j{(k)}=C_{j-1+k}^k.
\end{equation}
\end{definition}
By definition, $K_n^{(m)}(x)$ and $I_n^{(m)}(x)$ are related to $L_n^{(m)}(x)$ as
\begin{eqnarray}
K_n^{(m)}(x) & = & \frac{x}{1-x}L_n^{(m)}(x)+\frac{x}{n}\frac{d}{dx}L_n^{(m)}(x),\\
I_n^{(m)}(x) & = & L_n^{(m)}(x)-K_n^{(m)}(x).
\end{eqnarray}

\begin{lemma}
The function $L_n^{(m)}(x)$ has the following equivalent expressions
\begin{eqnarray}
L_n^{(m)}(x) & = & 1-nC_{n+m}^mx^{m+1}\sum_{l=0}^{n-1}C_{n-1}^l\frac{(-x)^l}{m+l+1},\label{eq:sum}\\
& = & 1-nC_{n+m}^m\int_0^x x^m(1-x)^{n-1}dx.\label{eq:integral}
\end{eqnarray}

\begin{proof}
In order to prove Eq. (\ref{eq:sum}), we decompose $L_n^{(m)}(x)$ into a polynomial $\sum_{k=0}^{n+m} c_kx^k$, and calculate the  coefficient $c_k$.\\
(i) For $k=0$, we have $c_0=f_n^{(0)}=1$.\\
(ii) For $1\le k\le m$, direct calculation leads to $c_k=\sum_{k'=0}^k (-1)^{k'}M_{k'}^{(k)}$, where $M_{k'}^{(k)}=\sum_{k'=0}^k C_{n}^{k'}f_n^{(k-k')}=\sum_{k'=0}^k C_{n}^{k'}C_{n+k-k'-1}^{k-k'}$. 
Further, by the definition of $M_{k'}^{(k)}$, the following equation holds for $k\geq1$ and $k'=1,\dots,k$,
\begin{eqnarray}
M_{k'}^{(k)}=\Delta_{k'-1}^{(k)}+\Delta_{k'}^{(k)},\label{eq:delta}
\end{eqnarray}
where $\Delta_0^{(k)}=C_n^0f_n^{(k)}=f_n^{(k)}$, and $\Delta_{k'}^{(k)}=M_{k'}^{(k)}\frac{(n-k')(k-k')}{nk}$.

Therefore, we get
\begin{eqnarray}
c_k & = & M_0+\sum_{k'=1}^k (-1)^{k'}(\Delta_{k'-1}+\Delta_{k'})\\
 & = & \sum_{k'=0}^k (-1)^{k'}\Delta_{k'}+\sum_{k'=0}^{k-1} (-1)^{k'+1}\Delta_{k'}\\
 & = & \Delta_{k}=0.
\end{eqnarray}
%Similarly , $\sum_{k'=0}^n (-1)^{k'}M_{k'}=0$ , for $k \ge n$.\\
(iii) For $m+1\le k\le m+n$, let $k=m+l$ and thus, $0\le l\le n-1$. The coefficient $c_k$ is calculated as
\begin{eqnarray}
c_k & = & \sum_{k'=l+1}^n (-1)^{k'}M_{k'}\\
 & = & \sum_{k'=0}^n (-1)^{k'}M_{k'}-\sum_{k'=0}^l (-1)^{k'}M_{k'}\\
 & = & 0-(-1)^{l}\Delta_{l}\\
 & = & (-1)^{l+1}\frac{n-l}{n}C_n^l\frac{n(m+1)}{n(m+l+1)}C_{n+m}^{m+1}\\
 & = & (-1)^{l+1}nC_{n+m}^{m}\frac{C_{n-1}^{l}}{m+l+1}.
\end{eqnarray}
This leads to the expression as Eq. (\ref{eq:sum}).

Further, from Eq. (\ref{eq:sum}) we have
\begin{eqnarray}
  L_n^{(m)}(0) & = & 1,\nonumber\\
  \frac{d}{dx}L_n^{(m)}(x) & = & -nC_{n+m}^m\sum_{l=0}^{n-1}C_{n-1}^l(-1)^lx^{m+l}\nonumber\\
  &=& -nC_{n+m}^m x^m (1-x)^{n-1}.
\end{eqnarray}
This leads to the expression as Eq. (\ref{eq:integral}).

% Furthermore,let
% \begin{equation}
% G(x) := (-1)^{m+1}nC_{n+m}^{m}\sum_{l=0}^{n-1}\frac{C_{n-1}^{l}(-x)^{m+l+1}}{m+l+1},\\\label{eq:oldg}
% \end{equation}
% then
% \begin{eqnarray}
% g(x) := \frac{d}{dx}G(x)& = & nC_{n+m}^{m}(-x)^{m}\sum_{l=0}^{n-1}C_{n-1}^l(-x)^{l}\\
% & = & nC_{n+m}^{m}(-x)^{m}(1-x)^{n-1}
% \end{eqnarray}
% So
% \begin{equation}
% G(x)=nC_{n+m}^{m}\int_0^x(-x)^{m}(1-x)^{n-1}dx\label{eq:newg}
% \end{equation}
% Replace Eq.(\ref{eq:oldg}) by  Eq.(\ref{eq:newg}) , Eq.(\ref{eq:sum}) turn into Eq.(\ref{eq:integral}).
\end{proof}
\end{lemma}

\begin{lemma}\label{lemma:In}
For $x\in(0,\frac12)$,
\begin{equation}
I_n^{(n-1)}(x) = \frac{1-2x}{1-x}+x\delta_n(1-x),
\end{equation}
with $\delta_n(1-x):=\sum_{n'=n}^\infty \frac{C_{2n'}^{n'}}{n'+1}[x(1-x)]^{n'}<o[(4x(1-x))^n n^{-3/2}]$.

\begin{proof}
From Eq.(\ref{eq:Ix}), $I^{(0)}_1(x)=1-x$, and
\begin{eqnarray}
\nonumber  I^{(n-1)}_{n}(x)&=&(1-x)^n[1+C_{n-1}^1x+\frac{1}{n}\sum_{j=2}^{n-1}(n-j)C_{n+j-1}^{j}x^j],\\
\nonumber &=&(1-x)^n[(1+x)^{n-1}+\frac{1}{n}\sum_{j=2}^{n-1}(n-j)C_{n+j-1}^{j}x^j-\sum_{j=2}^{n-1}C_{n-1}^{j}x^j],\\
%\nonumber &&+\frac{1}{n}\sum_{j=2}^{n-1}(n-j)C_{n+j-1}^{j}x^j-\sum_{j=2}^{n-1}C_{n-1}^{j}x^j],\\
  &=&(1-x)^n[(1+x)^{n-1}+\sum_{j=2}^{n-1}(\frac{n-j}{n}C_{n+j-1}^{j}-C_{n-1}^{j})x^j],\\
  %        &&-C_{n-1}^{j})x^j],\\
%\nonumber  I^{(n)}_{n+1}(x)&=&(1-x)^n[(1-x)(1+x)(1+x)^{n-1}\\
          I^{(n)}_{n+1}(x)&=&(1-x)^n[(1-x)(1+x)(1+x)^{n-1}+(1-x)\sum_{j=2}^{n}(\frac{n+1-j}{n+1}C_{n+j}^{j}-C_{n}^{j})x^j].
\end{eqnarray}
It follows that
\begin{eqnarray}
\nonumber  &&I^{(n)}_{n+1}(x)-I^{(n-1)}_{n}(x)\\
\nonumber &=&(1-x)^n \{-x^2(1+x)^{n-1}+\sum_{j=2}^{n-1}[(1-x)(\frac{n+1-j}{n+1}C_{n+j}^{j}-C_{n}^{j})\\
  \nonumber        && -(\frac{n-j}{n}C_{n+j-1}^{j}-C_{n-1}^{j})]x^j+(1-x)x^n(\frac{C^n_{2n}}{n+1}-1) \},\\
\nonumber  &=&x^2(1-x)^n\{-\sum^{n-1}_{j=0}C^j_{n-1}x^j+\sum_{j=0}^{n-3}[(1-x)(\frac{n-1-j}{n+1}C_{n+j+2}^{j+2}-C_{n}^{j+2})\\
 &&\nonumber -(\frac{n-2-j}{n}C_{n+j+1}^{j+2}-C_{n-1}^{j+2})]x^j+(1-x)x^{n-2}(\frac{C^n_{2n}}{n+1}-1) \},\\
 \nonumber  &=&x^2(1-x)^n\{\sum_{j=0}^{n-3}[-x(\frac{n-(j+1)}{n+1}C_{n+j+2}^{j+2}-C_{n}^{j+2})\\
 && \nonumber-(-\frac{n-j}{n+1}C_{n+j+1}^{j+1}+C^{j+1}_{n})]x^j-(n-1)x^{n-2}-x^{n-1}+(1-x)x^{n-2}(\frac{C^n_{2n}}{n+1}-1) \},\label{eq:sim1}\\
\nonumber  &=&x^2(1-x)^n\{\sum_{j=0}^{n-3}(C_{n}^{j+2}-\frac{n-(j+1)}{n+1}C_{n+j+2}^{j+2})x^{j+1}+\sum_{j=0}^{n-3}(\frac{n-j}{n+1}C_{n+j+1}^{j+1}-C^{j+1}_{n})x^j\\
          && -(n-1)x^{n-2}-x^{n-1}+(1-x)x^{n-2}(\frac{C^n_{2n}}{n+1}-1) \},\\    
\nonumber  &=&x^2(1-x)^n\{(n-\frac{2}{n+1}C^{n-1}_{2n-1})x^{n-2}-(n-1)x^{n-2}-x^{n-1}+(1-x)x^{n-2}(\frac{C^n_{2n}}{n+1}-1) \},\\
   \nonumber       &=&x^2(1-x)^n\{-x^{n-1}-x^{n-1}(\frac{C^n_{2n}}{n+1}-1) \},\\      
   &=&-x^{n+1}(1-x)^n\frac{C^n_{2n}}{n+1}.
\end{eqnarray}
In obtaining Eq.(\ref{eq:sim1}), we used the relations
$\frac{n-2-j}{n}C_{n+j+1}^{j+2}=\frac{n-1-j}{n+1}C_{n+j+2}^{j+2}-\frac{n-j}{n+1}C_{n+j+1}^{j+1}$ and 
$C_{n-1}^{j}-C_{n-1}^{j+2}=C_{n-1}^{j+1}+C_{n-1}^{j}-(C_{n-1}^{j+2}+C_{n-1}^{j+1})=C_{n}^{j+1}-C_{n}^{j+2}$.
Furthermore, for $x\in(0,\frac12)$, the following equation holds
\begin{equation}
 \sum^{\infty}_{n=1}\frac{C^{n}_{2n}}{n+1}[x(1-x)]^{n}=\frac{x}{1-x}.\label{eq:x1x}
\end{equation}
The proof is as follows. Let $x(1-x)=t$, and for $x\in(0,\frac12)$, we have $x=\frac{1-\sqrt{1-4t}}{2}$. Consequently, Eq. (\ref{eq:x1x}) becomes
\begin{equation}
 \sum^{\infty}_{n=1}\frac{C^{n}_{2n}}{n+1}t^{n}=\frac{1-2t-\sqrt{1-4t}}{2t}.\label{eq:x1x1}
\end{equation}

Taylor expansion of $\sqrt{1-4t}$ at $t=0$ gives\\
\begin{eqnarray}
\nonumber  \sqrt{1-4t}&=&1-2t-2\sum^{\infty}_{n=2}\frac{(2n-2)!}{n!(n-1)!}t^n,\\
\nonumber &=&1-2t-2\sum^{\infty}_{n=1}\frac{(2n)!}{n!(n+1)!}t^{n+1},\\
 &=&1-2t-2\sum^{\infty}_{n=1}\frac{C^n_{2n}}{n+1}t^{n+1},
\end{eqnarray}
which is equivalent to Eq. (\ref{eq:x1x1}), and thus we arrive at Eq. (\ref{eq:x1x}).

Therefore, for $x\in(0,\frac12)$,
\begin{eqnarray}
\nonumber  I^{(n-1)}_{n}(x)&=&I^{0}_1(x)+\sum^{n-1}_{n'=1}I^{(n)}_{n+1}(x)-I^{(n-1)}_{n}(x),\\
\nonumber &=&1-x-x\sum^{n-1}_{n'=1}\frac{C^{n'}_{2n'}}{n'+1}[x(1-x)]^{n'},\\
\nonumber &=&1-x-x(\frac{x}{1-x}-\sum^{\infty}_{n'=n}\frac{C^{n'}_{2n'}}{n'+1}[x(1-x)]^{n'}),\\
\nonumber &=&\frac{1-2x}{1-x}+x\delta_n(1-x).\\
\end{eqnarray}
For $n\gg 0$, by Stirling's approximation $n!\approx\sqrt{2\pi n}(\frac{n}{e})^n$, we have
\begin{eqnarray}
\nonumber  \delta_n(1-x)&=&\sum_{n'=n}^\infty \frac{C_{2n'}^{n'}}{n'+1}[x(1-x)]^{n'},\\
\nonumber  &\approx&\sum_{n'=n}^\infty \frac{\sqrt{2\pi(2n')}}{2\pi n'}\frac{(\frac{2n'}{e})^{2n'}}{(\frac{n'}{e})^{2n'}}\frac{1}{n'+1}[x(1-x)]^{n'},\\
\nonumber &=&\sum_{n'=n}^\infty \frac{1}{\sqrt{\pi n'}(n'+1)}[4x(1-x)]^{n'},\\
\nonumber &<&\sum_{n'=n}^\infty \frac{1}{\sqrt{\pi n}(n+1)}[4x(1-x)]^{n'}\\
\nonumber &<&\frac{1}{\sqrt{\pi }n^{3/2}}\frac{[4x(1-x)]^{n}}{(2x-1)^2}
=o[(4x(1-x))^n n^{-3/2}].
\end{eqnarray}
% \begin{equation}
% \delta_n(1-x)=\frac{[4x(1-x)]^{n}}{(2x-1)^2}[\frac{1}{(n+1)\sqrt{n\pi }}+o(n^{-\frac{3}{2}})]
% \end{equation} 
\end{proof}
\end{lemma}

\begin{lemma} \label{lemma: Lx}
In the limit $n \to \infty$, the function $L_n^{(m)}(x)$ approaches a discontinuous function as
\begin{equation}
 \lim_{n \to \infty}  L_n^{(m)}(x)=
% L_n^{(m)}(x)=
\left\{\begin{matrix}
 1,    &x <\frac{m}{m+n-1}, \\
  0,   & x > \frac{m}{m+n-1}.
\end{matrix}\right.\label{eq:limit}
\end{equation} 

\begin{proof}
By definition, we have $L_n^{(m)}(0)=1$ and $L_n^{(m)}(1)=0$, $\forall m,n$.

Now let $g^{(m)}_n(x):=-\frac{d}{dx}L_n^{(m)}(x)=nC_{n+m}^{m}x^{m}(1-x)^{n-1}$.

For $n \gg 0$ and $m$ being finite, or equivalently, $\frac{m}{m+n-1}\rightarrow 0_+$, we have
\begin{eqnarray}
\nonumber  g^{(m)}_n(x)&=&\frac{n}{m!}\frac{(n+m)!}{n!}x^{m}(1-x)^{n-1},\\
\nonumber & \approx &\frac{n}{m!}\sqrt{\frac{n+m}{n}} \frac{(\frac{n+m}{e})^{m+n}}{(\frac{n}{e})^{n}}x^{m}(1-x)^{n-1},\\
\nonumber &=&\frac{n}{m!}\sqrt{\frac{n+m}{n}}(\frac{n+m}{e})^{m}(\frac{n+m}{n})^{n}x^{m}(1-x)^{n-1},\\
\nonumber & \approx &\frac{n}{m!}\sqrt{\frac{n+m}{n}}(n+m)^{m}x^{m}(1-x)^{n-1}.
%\nonumber & \to &n^{m+1}(1-x)^{n-1} \to 0,\\
\end{eqnarray}
In the second line, we used Stirling's formula, and in the fourth line, we used $(\frac{n+m}{n})^{n} \approx e^m$ for $n \gg 0$. The above equation leads to $\lim_{n\rightarrow\infty} g^{(m)}_n(x)=0$, $\forall x\in (0,1]$. Therefore, 
\begin{equation}
\lim_{n \to \infty} L_n^{(m)}(x)=
\left\{\begin{matrix}
 1,    &x =0, \\
  0,   & 0<x \le 1.
\end{matrix}\right.
\end{equation} 
It means that Eq. (\ref{eq:limit}) holds for $\frac{m}{m+n-1}\rightarrow 0_+$.

For $n\gg0$ and $\frac{m}{m+n-1}\neq 0$, which means that $m\gg0$,
\begin{eqnarray}
\nonumber  g^{(m)}_n(x)&=&(n+m)\frac{(n-1+m)!}{m!(n-1)!}x^{m}(1-x)^{n-1},\\
\nonumber & \approx &\frac{n+m}{\sqrt{2\pi}}\sqrt{\frac{n-1+m}{m(n-1)}}\frac{(n-1+m)^{n-1+m}}{m^m(n-1)^{n-1}}x^{m}(1-x)^{n-1},\\
\nonumber &=&\frac{n+m}{\sqrt{2\pi}}\sqrt{\frac{n-1+m}{m(n-1)}}[\frac{n-1+m}{m}x]^m[\frac{n-1+m}{n-1}(1-x)]^{n-1},\\
\nonumber &=&\frac{n+m}{\sqrt{2\pi}}\sqrt{\frac{n-1+m}{m(n-1)}}[(1+\frac{1}{\mu})^\mu x^\mu(1+\mu)(1-x)]^{n-1}.\\
\end{eqnarray}
where $\mu=\frac{m}{n-1} \le 1$. 
Notice that $x^\mu(1-x) \le [(1+\frac{1}{\mu})^\mu(1+\mu)]^{-1}$, and the equation holds if and only if $x=\frac{\mu}{1+\mu}=\frac{m}{m+n-1}$. Hence,
\begin{equation}
 \lim_{n \to \infty}  -\frac{d}{dx}L_n^{(m)}(x)=nC_{n+m}^{m}x^{m}(1-x)^{n-1}=
 %g_n^{(m)}(x) \stackrel{n \to \infty}{\longmapsto} 
\left\{\begin{matrix}
 \infty,    & x=\frac{m}{m+n-1} ,\\
  0,   & otherwise.
\end{matrix}\right.\label{eq:limitg}
\end{equation} 
Combining it with the fact that $L_n^{(m)}(0)=1$ and $L_n^{(m)}(1)=0$, we arrive at Eq. (\ref{eq:limit}).
%In other words, for $x \ne \frac{m}{m+n-1}$, $K_n^{(m)}(x)=\frac{x}{1-x}L_n^{(m)}(x)$.
\end{proof}
\end{lemma}

\begin{definition}
    The function $I_d(x,y)$ with $d$ being a positive integer and $x,y\in[0,1]$ is defined as
\begin{equation}
I_d(x,y):=\frac{(1-x)^d(1-y)^d}{d}\sum^{d-1}_{j=0}\sum^{d-j-1}_{k=0}(d-j-k)f^{(k)}_{d}x^{k}f^{(j)}_{d}y^j.
\end{equation} 
\end{definition}

\begin{lemma}
    In the limit $d\rightarrow\infty$, the function $I_d(x,y)$ becomes
\begin{equation}\label{eq:Id}
 \lim_{d \to \infty}  I_d(x,y)=
\left\{\begin{matrix}
  1-\frac{x}{1-x}-\frac{y}{1-y}, & \frac{y}{1-y} < 1-\frac{x}{1-x}, \\
 0,   & \frac{y}{1-y} \ge 1-\frac{x}{1-x}.
\end{matrix}\right.
\end{equation} 

\begin{proof}
    Let $r_d^{(j)}(x):=\frac{(1-x)^d}{d}\sum^{d-j-1}_{k=0}(d-j-k)f^{(k)}_{d}x^{k}$, and then, 
    \begin{equation}
        I_d(x,y)=(1-y)^d\sum^{d-1}_{j=0}r_d^{(j)}(x)f^{(j)}_{d}y^j.\label{eq:app_Id1}
    \end{equation}

By definition, the function $r_d^{(j)}(x)$ is calculated as
\begin{eqnarray}
\nonumber  r_d^{(j)}(x)  &=&\frac{d-j}{d}L^{(d-j-1)}_d(x)-K^{(d-j-1)}_d(x),\\
 &=&\big(\frac{d-j}{d}-\frac{x}{1-x} \big)L^{(d-j-1)}_d(x).
\end{eqnarray}
In the limit of large $d$, by Lemma \ref{lemma: Lx}, we have
\begin{equation}
\lim_{d \to \infty} r_d^{(j)}(x)=
\left\{\begin{matrix}
  \frac{d-j}{d}-\frac{x}{1-x},    &0 \le j < j_x ,\\
  0,   &  j_x < j \le d-1.
\end{matrix}\right.
\end{equation} 
where $j_x=\frac{(d-1)(1-2x)}{1-x}$.

Submitting the above expression for $r_d^{(j)}(x)$ into Eq. (\ref{eq:app_Id1}), we have for $d\rightarrow\infty$,
\begin{eqnarray}
\nonumber I_d(x,y)&=&(1-y)^d\sum^{d-1}_{j=0}\big(\frac{d-j}{d}-\frac{x}{1-x} \big)L^{(d-j-1)}_d(x)f^{(j)}_{d}y^j,\\
\nonumber   &=&(1-y)^d\sum^{\left \lfloor j_x \right \rfloor }_{j=0}\big(\frac{d-j}{d}-\frac{x}{1-x} \big)f^{(j)}_{d}y^j,\\
\nonumber   &=&(1-\frac{x}{1-x})L^{(\left \lfloor j_x \right \rfloor )}_d(y)-K^{(\left \lfloor j_x \right \rfloor )}_d(y),\\
\nonumber   &=&(1-\frac{x}{1-x}-\frac{y}{1-y})L^{(\left \lfloor j_x \right \rfloor )}_d(y).\\
\end{eqnarray}
%for $\forall x,y \in (0,1)$.
For $y < \frac{\left \lfloor j_x \right \rfloor}{\left \lfloor j_x \right \rfloor+d-1}$, or equivalently, $\frac{y}{1-y} < 1-\frac{x}{1-x}$, $L^{(\left \lfloor j_x \right \rfloor)}_d(y)=1$, and otherwise, $L^{(\left \lfloor j_x \right \rfloor)}_d(y)=0$. Therefore, we arrive at Eq. (\ref{eq:Id}).
\end{proof}
\end{lemma}

\section{QUBIT STATE TRANSFORMATIONS UNDER \texorpdfstring {$\bm{\mmtpd}$}{}} \label{sec:qubit_app}
\subsection{A motivating example}
Here we consider a simplest example, where the target system $S$ is a qubit with Hamiltonian $H_S=E|e\rangle\langle e|$ and initially in state $[p_0,1-p_0]$, and the memory is a qubit with Hamiltonian $H_A=0$ and initially in state $[\frac12,\frac12]$.
The proposal for attaining the ground population $p^{(2)}$ as in Eq. (\ref{eq:p2}) goes as follows.
The initial state of $SA$ can be expressed as a probability distribution on basis $\{|g0\rangle, |g1\rangle, |e0\rangle, |e1\rangle\}$, i.e.,
\begin{equation}
\boldsymbol{p}=\frac12[p_0,p_0;1-p_0,1-p_0].
\end{equation}
The global MTP consist of two steps. In the first step, full elementary thermalizations are implemented subsequently between $|g0\rangle$ and each of $|ej\rangle$ ($j=1,2$), and the state becomes
\begin{eqnarray}
\boldsymbol{p_1}=T^{g0,e1}T^{g0,e0}\boldsymbol{p}=\frac12[\gamma(1-p_0+\gamma),p_0;1-\gamma,(1-\gamma)(1-p_0+\gamma)].
\end{eqnarray}
Similarly, the second step consists of $T^{g1,e0}$ and $T^{g1,e1}$ in turn, and output state reads
\begin{eqnarray}
\boldsymbol{p_2}=T^{g1,e1}T^{g1,e0}\boldsymbol{p_1}=[a_2^{(1)},a_2^{(2)};b_1^{(2)},b_2^{(2)}],
\end{eqnarray}
with
\begin{eqnarray}
a_2^{(1)}&=&\frac12\gamma(1-p_0+\gamma),\nonumber\\
a_2^{(2)}&=&\frac12\gamma[1+(2\gamma-1)(p_0-\gamma)],\nonumber\\
b_1^{(2)}&=&\frac12(1-\gamma)[p_0+1-\gamma],\nonumber\\
b_2^{(2)}&=&\frac12(1-\gamma)[1+(2\gamma-1)(p_0-\gamma)].\label{eq:step1}
\end{eqnarray}
Then, full thermalization is applied to the auxiliary system $A$, leaving the target state in $\rho^{(2)}$ with $p^{(2)}=a_2^{(1)}+a_2^{(2)}$, which equals to Eq. (\ref{eq:p2}).

\subsection{Simulation of \texorpdfstring {$\beta$}{}-swap by \texorpdfstring {$\bm{\mmtpd}$}{}} \label{sec:nd}
We propose a proposal for simulating $\beta$-swap via $\mmtpd$. Our results show that, for $d$ large enough, we can use $\mmtpd$ to approximate any qubit state transformation enabled by TP. Our proposal consists of $d^2$ elementary full thermalizations and a full thermalization locally acting on $A$ as a last step. Precisely, the protocol goes as follows.

Initially, the state of $SA$ is prepared as
\begin{equation}
\rho=\big[ p_0|g\rangle\langle g|+(1-p_0)|e\rangle\langle e| \big]\otimes\frac1d\sum_{l=1}^{d}|l\rangle\langle l|.
\end{equation}
On basis $\{|g1\rangle,\dots,|gd\rangle,|e1\rangle,\dots,|ed\rangle\}$, the probability distribution vector of the initial state is written as
\begin{equation}
\boldsymbol{p_{0d}}=[a_0^{(1)},\dots,a_0^{(d)};b_1^{(0)},\dots,b_d^{(0)}],
\end{equation}
with
\begin{equation}
a_0^{(k)}=\frac{p_0}{d},\ b_j^{(0)}=\frac{1-p_0}{d}.
\end{equation}
In the $k$th step ($k=1,\dots,d$), full elementary thermalizations are implemented subsequently between $|gk\rangle$ and each of $|ej\rangle$ ($j=1,\dots,d$). Then we have
\begin{equation}
\boldsymbol{p_{kj}}=\big\{
\begin{array}{cc}
T^{gk,e1}\boldsymbol{p_{k-1,d}}, & j=1,\\
T^{gk,ej}\boldsymbol{p_{k,j-1}}, & 2\leq j\leq d,
\end{array}
\end{equation} \label{eq:pkj}
where
\begin{eqnarray}
\boldsymbol{p_{kj}}=[ a_d^{(1)},\dots,a_d^{(k-1)},a_j^{(k)},a_0^{(k+1)},\cdots,a_0^{(d)}; b_1^{(k)},\dots, b_j^{(k)}, b_{j+1}^{(k-1)}, \dots, b_d^{(k-1)}].
\end{eqnarray}
Direct calculation leads to
\begin{eqnarray}\label{eq:ajk_bjk}
a_j^{(k)} = \gamma (a_{j-1}^{(k)}+b_j^{(k-1)}),
b_j^{(k)} = (1-\gamma) (a_{j-1}^{(k)}+b_j^{(k-1)}).
\end{eqnarray}
Therefore, we have
\begin{equation}\label{eq:bjk}
    b_j^{(k)}=\frac{1-\gamma}{\gamma}a_j^{(k)},
\end{equation}
and Eq. (\ref{eq:ajk_bjk}) becomes
\begin{eqnarray}
a_j^{(k)} & = & \gamma a_{j-1}^{(k)} + (1-\gamma)a_j^{(k-1)},\nonumber\\
a_0^{(k)} & = & \frac{p_0}{d},\ a_j^{(0)}=\frac{\gamma}{1-\gamma}b_j^{(0)}=\frac{\gamma}{1-\gamma}\frac{1-p_0}{d}.
\end{eqnarray}
Now we make the substitution 
\begin{equation}\label{eq:ajk}
    a_j^{(k)}=\frac1d\gamma^j(1-\gamma)^{k-1}s_j^{(k)},
\end{equation}
and the above problem is equivalent to
\begin{eqnarray}
s_j^{(k)} & = & s_{j-1}^{(k)} + s_j^{(k-1)},\nonumber\\
s_0^{(k)} & = & p_0(1-\gamma)^{-k+1},\ s_j^{(0)}=(1-p_0)\gamma^{-j+1}.\label{eq:s}
\end{eqnarray}
This is in turn equivalent to
\begin{equation}
s_j^{(k)}=(1-\gamma)^{-k}[(1-p_0)\gamma^{-j+1}-(\gamma-p_0)\sum_{k'=0}^{k-1}f_j^{(k')}(1-\gamma)^{k'}],\label{eq:sjk}
\end{equation}
where $f_j^{(k)}:=\sum_{j'=1}^j f_{j'}^{(k-1)},\ f_j^{(0)}=1$, or equivalently, $f_j^{(k)}=C_{j+k-1}^k$.
Here $C_n^m:=\frac{n!}{m!(n-m)!}$.
The proof goes as follows.
For $k=0$, Eq. (\ref{eq:sjk}) reduces to $s_j^{(0)}=(1-p_0)\gamma^{-j+1}$, which is equivalent to Eq. (\ref{eq:s}).
It means that Eq. (\ref{eq:sjk}) holds for the case with $k=0$ and $j=1,\dots,d$.
Now assume it holds for the case with $k-1$ and $j=1,\dots,d$, and we will prove it also holds for $k$.
From Eq. (\ref{eq:s}), we have $s_j^{(k)}-s_{j-1}^{(k)} = s_j^{(k-1)}$. Taking the summation over all $j$ gives
\begin{equation}
\sum_{j'=1}^d[s_j^{(k)}-s_{j-1}^{(k)}]=\sum_{j'=1}^d s_j^{(k-1)}.
\end{equation}
It follows that
 \begin{eqnarray}%\label{eq:sjk}
s_{j}^{(k)}&=& s_0^{(k)}+(1-\gamma)^{1-k}\big[(1-p_0)\sum_{j'=1}^j\gamma^{1-j'}-(\gamma-p_0)\sum_{k'=0}^{k-2}\sum_{j'=1}^jf_{j'}^{(k')}(1-\gamma)^{k'}\big]\nonumber\\
&=& s_0^{(k)}+(1-\gamma)^{1-k}\big[(1-p_0)\frac{1-\gamma^{-j}}{1-\gamma^{-1}}-(\gamma-p_0)\sum_{k'=1}^{k-1}f_{j'}^{(k')}(1-\gamma)^{k'-1}\big]\nonumber\\
&=& (1-\gamma)^{-k}[(1-p_0)\gamma^{-j+1}-(\gamma-p_0)\sum_{k'=0}^{k-1}f_j^{(k')}(1-\gamma)^{k'}].
\end{eqnarray}
In the last line, we used Eq. (\ref{eq:s}).

After the $d^2$ elementary full thermalizations, the state becomes
\begin{eqnarray}
\boldsymbol{p_{dd}}=[  a_d^{(1)},\dots,a_d^{(d)};b_1^{(d)},\dots, b_d^{(d)}].
\end{eqnarray}
It follows from Eq. (\ref{eq:sjk}) that
\begin{equation}
a_d^{(k)}=\frac1d\frac{\gamma}{1-\gamma}(1-p_0)-\frac{\gamma-p_0}{d(1-\gamma)}\gamma^d\sum_{k'=0}^{k-1}f_d^{(k')}(1-\gamma)^{k'}.\label{eq:adk}
\end{equation}

Then full thermalization is implemented locally on $A$, and the state of the qubit becomes $[p^{(d)},1-p^{(d)}]$ with
$p^{(d)}\equiv\sum_{k=1}^{d}a_d^{(k)}$. Direct calculations lead to
\begin{equation}
p^{(d)}=\frac{\gamma}{1-\gamma}(1-p_0)-\frac{\gamma-p_0}{1-\gamma}I_d^{(d-1)}(1-\gamma),
\end{equation}
where the function $I_d^{(d-1)}(x)$ with $x\in(0,1)$ is defined in Appendix \ref{sec:app1}. It decreases with $d$ and for finite $d$ we have
$I_d^{(d-1)}(1-\gamma)\in(\frac{2\gamma-1}{\gamma},\gamma]$. From Lemma \ref{lemma:In} in Appendix \ref{sec:app1}, we have
\begin{equation}
p^{(d)}=1-p_0e^{-\beta E}-(\gamma-p_0)\delta_d(\gamma),\label{eq:pd}
\end{equation}
where $\delta_d(\gamma)<o[(4\gamma(1-\gamma))^d d^{-3/2}]$. Therefore, $\beta$-swap can be simulated by our protocol with error exponentially decreasing with the dimension of the assisted memory.

\section{Three-dimension target system and the comparision between \texorpdfstring {$\bm{\mmtpd}$}{} and ETP}\label{sec:3D}

When the dimension of target system is more than 2, ETP is a strict subset of TP. Therefore, it is of interest to investigate the comparison between $\mmtpd$ and ETP, especially when $d$ is not large.

Consider a three-dimensional system with Hamiltonian $H_S=0|g\rangle\langle g|+E(|e_1\rangle\langle e_1|+|e_2\rangle\langle e_2|)$ and initially in the ground state $|g\rangle$. The occupation probability vector of the initial state reads $[1,0,0]$.

The boundaries of states that can be obtained from $|g\rangle$ via TP, ETP, as well as MTP, are plotted in Fig. \ref{fig:3Dcone} .
Now we show that with the assistance of a qubit memory, MTP can achieve state transformation that cannot be achieved by ETP.

The vertices can be obtained as\\
%\begin{widetext}
Vertex $A_s$($s=1,2$):$T_MT^{g2,e_s2}T^{g2,e_s1}T^{g1,e_s2}T^{g1,e_s1}$.\\
Vertex $B_s$ ($s=1,2$):$T_MT^{g2,e_{\bar{s}}2}T^{g2,e_{\bar{s}}1}T^{g1,e_{\bar{s}}2}T^{g1,e_{\bar{s}}1}T^{g2,e_s2}T^{g2,e_s1}T^{g1,e_s2}T^{g1,e_s1}$.\\
%\end{widetext}
Other states on the boundary can be obtained by elementary thermalizations from the vertices. The states inside can be obtained by thermalization from boundary states.

\begin{figure}[htbp] 
\centering 
\includegraphics[width=0.5\textwidth]{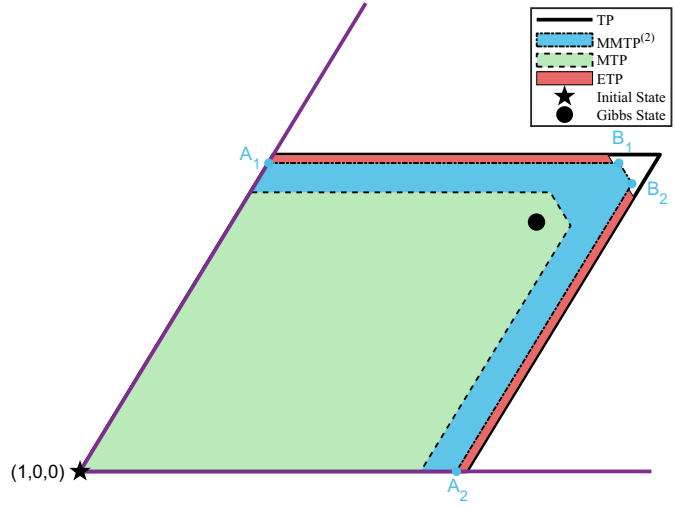} 
\caption{\textbf{Comparison of the sets of state that can be reached via TP, ETP, MTP,and \texorpdfstring {$\bm{\mmtpt}$}{} from a qutrit state.}
The Hamiltonian of the qutrit reads $H_S=0|g\rangle\langle g|+E(|e_1\rangle\langle e_1|+|e_2\rangle\langle e_2|)$, and the qutrit is initially in its ground state $|g\rangle$. This example shows that via $\mmtpt$ one can realize state transformations that cannot be achieved by ETP.} 
\label{fig:3Dcone} 
\end{figure}

\section{Details in the memory-assisted protocol for work extraction}\label{sec:proof}
The protocol is summarized in the following scheme
\begin{eqnarray}
  \nonumber &&[0,0,1,0]\otimes[\overbrace{\frac{1}{d},...,\frac{1}{d}}^{d \ times}] \\
 % \nonumber && =[\overbrace{0,...,0}^d,\overbrace{0,...,0}^d,\overbrace{\frac{1}{d},...,\frac{1}{d}}^d,\overbrace{0,...,0}^d]\\
   \nonumber && =[{0,...,0};0,...,0;\frac{1}{d},...,\frac{1}{d};0,...,0]\\
  \nonumber && \stackrel{\textbf{(\Rmnum{1})}}{\longmapsto} [0,...,0;a_d^{(1)},...,a_d^{(d)};b_1^{(d)},...,b_d^{(d)};0,...,0]\\
\nonumber  &&\stackrel{\textbf{(\Rmnum{2})}}{\longmapsto} [0,...,0;a_d^{(1)},...,a_d^{(d)};\epsilon_1,...,\epsilon_d;c_1^{(d)},...,c_d^{(d)}].
\end{eqnarray}

Step \Rmnum{1} consists of $d$ subroutines. In the $j$th subroutine, full elementary thermalizations are consequently implemented between $|e0j\rangle$ and each of $|g1k\rangle$ with $k=1,\dots,d$. This is equivalent to simulated $\beta$-swap between $|e0\rangle_{SW}$ and $|g1\rangle_{SW}$. By putting Eqs. (\ref{eq:bjk}), (\ref{eq:ajk}) and (\ref{eq:sjk}) together, we obtain
\begin{equation}
    b_j^{(d)}=\frac1d[1-\gamma_\delta^j\sum^{d-1}_{k=0}f^{(k)}_j(1-\gamma_\delta)^{k}].
\end{equation}
Further, because
\begin{eqnarray}
\nonumber  \sum^{j-1}_{j'=0}f^{(k)}_{j'+1}\gamma_\delta^{j'}-f^{(k+1)}_j\gamma_\delta^j &=& f^{(k+1)}_1+\sum^{j-1}_{j'=1}(f^{(k+1)}_{j'+1}-f^{(k+1)}_{j'})\gamma_\delta^{j'}-f^{(k+1)}_j\gamma_\delta^j
=\sum^{j-1}_{j'=0}f^{(k+1)}_{j'+1}\gamma_\delta^{j'}-\sum^{j-1}_{j'=0}f^{(k+1)}_{j'+1}\gamma_\delta^{j'+1}\\
  &=&(1-\gamma_\delta)\sum^{j-1}_{j'=0}f^{(k+1)}_{j'+1}\gamma_\delta^{j'},
\end{eqnarray}
we arrive at
\begin{eqnarray}
\nonumber b^{(d)}_j
 &=&\frac{1}{d}[(1-\gamma_\delta^j)-\gamma_\delta^j\sum^{d-1}_{k=1}f^{(k)}_j(1-\gamma_\delta)^{k}]
 = \frac{1-\gamma_\delta}{d}[\sum^{j-1}_{j'=0}f^{(0)}_{j'+1}\gamma_\delta^{j'}-f^{(1)}_j\gamma_\delta^j-\gamma_\delta^j\sum^{d-1}_{k=2}f^{(k)}_j(1-\gamma_\delta)^{k-1}]\\
\nonumber  &=&\frac{(1-\gamma_\delta)^2}{d}[\sum^{j-1}_{j'=0}f^{(1)}_{j'+1}\gamma_\delta^{j'}-f^{(2)}_j\gamma_\delta^j-\gamma_\delta^j\sum^{d-1}_{k=3}f^{(k)}_j(1-\gamma_\delta)^{k-2}]
 = ...
 =\frac{(1-\gamma_\delta)^{d-1}}{d}[\sum^{j-1}_{j'=0}f^{(d-2)}_{j'+1}\gamma_\delta^{j'}-f^{(d-1)}_j\gamma_\delta^j]\\
  &=&\frac{(1-\gamma_\delta)^d}{d}\sum^{j-1}_{j'=0}f^{(d-1)}_{j'+1}\gamma_\delta^{j'}.
\end{eqnarray}
Notice that $b^{(d)}_j$ is increasing in $j$.

Step \Rmnum{2} also consists of $d$ subroutines. 
In the following, we will label the action of Step (\Rmnum{2}) as $\hat\mT$, and thus,
\begin{equation}\label{eq:hatmT}
    [b_1^{(d)},...,b_d^{(d)};0,...,0]\stackrel{\hat\mT}{\longmapsto} [\epsilon_1,...,\epsilon_d;c_1^{(d)},...,c_d^{(d)}].
\end{equation}
Each subroutine of $\hat\mT$ is realized as $\hat\mT^{(k)}\equiv T^{e0k,e1d}\dots T^{e0k,e11}$.
% In each subroutine, full elementary thermalizations are subsequentially implemented between each one of $|e0k\rangle$ and $|e1j\rangle$ with $j=1,\dots,d$. 
In order to decide the order of $\hat\mT^{(k)}$ in $\hat\mT$, we first consider the following problem. 

Let $H_{eff}=0|\alpha\rangle\langle \alpha|+W\sum_{j=1}^d|\beta j\rangle\langle \beta j|$ be the effective Hamiltonian on a $(d+1)$-dimensional subspace spanned by $\{|\alpha\rangle;|\beta1\rangle,\dots,|\beta d\rangle\}$, and $\mT_{eff}=T^{\alpha,\beta d}\dots T^{\alpha,\beta 2} T^{\alpha,\beta 1}$ be a Markovian thermal process acting on this subspace. We define the parameters $d_k$ and $c_j^{(k)}$ ($j,k=1,\dots,d$) as
\begin{eqnarray}
    &[1;0,\dots,0]&\stackrel{\mT_{eff}}{\longmapsto} [d_1;c_1^{(1)},\dots,c_d^{(1)}],\label{eq:mTeff1}\\
    &[0;c_1^{(k-1)},\dots,c_d^{(k-1)}]&\stackrel{\mT_{eff}}{\longmapsto} [d_k;c_1^{(k)},\dots,c_d^{(k)}],\ k=2,\dots,d.\label{eq:mTeff2}
\end{eqnarray}
Then $d_k$ is calculated as follows. Because 
\begin{eqnarray}
  \nonumber &&[1;\overbrace{0,...,0}^{d\ times}] 
  \stackrel{T^{\alpha,\beta 1}}{\longmapsto} [\gamma_{_\W};1-\gamma_{_\W},\overbrace{0,...,0}^{d-1\ times}]
  \stackrel{T^{\alpha,\beta 2}}{\longmapsto} [\gamma_{_\W}^2;1-\gamma_{_\W},\gamma_{_\W}(1-\gamma_{_\W}),\overbrace{0,...,0}^{d-2\ times}]
   \  \stackrel{}{\longmapsto}\dots\\
  \nonumber && \stackrel{T^{\alpha,\beta d}}{\longmapsto} [\gamma_{_\W}^d;1-\gamma_{_\W},\gamma_{_\W}(1-\gamma_{_\W}),...,(1-\gamma_{_\W})\gamma_{_\W}^{d-1}],
\end{eqnarray}
from Eq. (\ref{eq:mTeff1}), we have $d_1=\gamma_{_\W}^d$, $c^{(1)}_j=(1-\gamma_{_\W})\gamma_{_\W}^{j-1}$.
Similarly, from Eq. (\ref{eq:mTeff2}), we have
\begin{eqnarray}
d_{k+1}&=&\gamma_{_\W}^{d}c_1^{(k)}+\gamma_{_\W}^{d-1}c_2^{(k)}+...+\gamma_{_\W} c_d^{(k)},\\
c^{(k+1)}_j&=&(1-\gamma_{_\W})(\gamma^{j-1}_{_\W}c^{(k)}_1+\gamma^{j-2}_{_\W}c^{(k)}_2+...+c^{(k)}_j).
\end{eqnarray}
Now let $c^{(k)}_j=(1-\gamma_{_\W})^k\gamma_{_\W}^{j-1}\chi_{j}^{k}$, and we have $\chi_j^{k+1}=\sum^j_{j'=1}\chi_{j'}^{k}$, and
$c^{(1)}_j=(1-\gamma_{_\W})\gamma_{_\W}^{j-1}=(1-\gamma_{_\W})\gamma_{_\W}^{j-1}\chi_j^{1}$.
Hence, $\chi_{j}^k=f_{j}^{(k-1)}$,$c^{(k)}_j=(1-\gamma_{_\W})^k\gamma_{_\W}^{j-1}f_{j}^{(k-1)}$. In turn, we get
$d_{k+1}=\sum^d_{j=1}\gamma_{_\W}^{d+1-j}c^{(k)}_j=\gamma_{_\W}^d(1-\gamma_{_\W})^kf^{k}_d$. Therefore, we arrive at
\begin{equation}\label{eq:dk}
    d_k=\gamma_{_\W}^d(1-\gamma_{_\W})^{k-1}f^{(k-1)}_d.
\end{equation}

Now consider a $2d$-subspace spanned by $\{|\alpha1\rangle,\dots,|\alpha d\rangle;|\beta 1\rangle,\dots,|\beta d\rangle\}$, and let $\vec x=[x_1,x_2,...,x_d;0,...,0]$ be a probability distribution in this subspace. (Notice that we do $not$ require $\sum_{j=1}^dx_j=1$.) Here $|\alpha j\rangle$ and $|\beta k\rangle$ are energy eigenstates with energy $E_{\alpha}$ and $E_{\beta}$, respectively, which satisfy $E_{\beta}-E_{\alpha}=W$.
We label $\mT=\mT^{(d)}\dots\mT^{(1)}$, where $\mT^{(k)}=T^{\alpha k,\beta d}\dots T^{\alpha k,\beta1}$, and calculate the output probability distribution after the action of $\mT$ on $\vec x$. 

The operator $\mT^{(1)}$ acts on the subspace spanned by $\{|\alpha1\rangle;|\beta1\rangle,\dots,|\beta d\rangle\}$. In this subspace, the action of $\mT^{(1)}$ is effectively the action of $\mT_{eff}$, so the state transforms as
\begin{equation}
    x_1[1;0,...,0]\stackrel{\mT^{(1)}}{\longmapsto} x_1[d_1;c_1^{(1)},...,c_d^{(1)}].
\end{equation}
%where the explicit forms of $d_k$ and $c_j^{(k)}$ are to be derived in \ref{app:sub_dk}.
Similarly, $\mT^{(2)}$ acts on the subspace spanned by $\{|\alpha2\rangle;|\beta1\rangle,\dots,|\beta d\rangle\}$, and also effectively equivalent to $\mT_{eff}$. The input state of $\mT^{(2)}$ in this subspace reads
\begin{equation}
    [x_2;x_1c_1^{(1)},...,x_1c_d^{(1)}]=x_2[1;0,...,0]+x_1[0;c_1^{(1)},...,c_d^{(1)}].
\end{equation}
It follows that
\begin{equation}
    [x_2;x_1c_1^{(1)},...,x_1c_d^{(1)}]\stackrel{\mT^{(2)}}{\longmapsto} x_2[d_1;c_1^{(1)},...,c_d^{(1)}]+x_1[d_2;c_1^{(2)},...,c_d^{(2)}].
\end{equation}

Analogously, $\mT^{(k)}$ acts on the subspace spanned by $\{|\alpha k\rangle;|\beta1\rangle,\dots,|\beta d\rangle\}$. The input state of $\mT^{(k)}$ in this subspace reads
\begin{equation}
    x_k[1;0,...,0]+x_{k-1}[0;c_1^{(1)},...,c_d^{(1)}]+\dots+x_1[0;c_1^{(k-1)},...,c_d^{(k-1)}].
\end{equation}
After the action of $\mT^{(k)}$, the above state becomes
\begin{equation}
    x_k[d_1;c_1^{(1)},...,c_d^{(1)}]+x_{k-1}[d_1;c_1^{(1)},...,c_d^{(1)}]+\dots+x_1[d_{k};c_1^{(k)},...,c_d^{(k)}].
\end{equation}
Therefore, the population on $|\alpha k\rangle$ after the action of $\mT^{(k)}$ reads
\begin{equation}
    \epsilon_k=d_1x_k+d_2x_{k-1}+...+d_kx_1,
\end{equation}
which preserves during the actions of subsequent operations $\mT^{(k')}$ with $k'>k$. Finally, the total population on $|\alpha1\rangle,\dots,|\alpha d\rangle$ reads
\begin{eqnarray}
\epsilon^{(d)}&=&\sum^d_{k=1}\epsilon_k
=d_{1}x_d+...+\sum_{k=1}^{d+1-i}d_{k}x_i+...+\sum_{k=1}^{d}d_{k}x_1
=\sum_{i=1}^d\left(\sum_{k=1}^{d+1-i}d_{k}\right)x_i.
\end{eqnarray}
Because $d_k\geq 0$, $\epsilon^{(d)}$ is minimized when $x_i$ is increasing in $i$. Therefore, we choose $b_j^{(d)}=x_j$ in the original problem as in Eq. (\ref{eq:hatmT}), and the optimal order of $\hat\mT^{(k)}$ is $\hat\mT=\hat\mT^{(d)}\dots\hat\mT^{(2)}\hat\mT^{(1)}$.

Finally, the error of work extraction under our memory-assisted protocol is calculated as
\begin{eqnarray}
 \nonumber  \epsilon^{(d)}&=&\sum^{d}_{j=1}\sum^{d-j+1}_{k=1}d_kb^{(d)}_j
 =\sum^{d}_{j=1}\sum^{d-j+1}_{k=1}\gamma_{_\W}^d(1-\gamma_{_\W})^{k-1}f^{(k-1)}_d\frac{(1-\gamma_\delta)^d}{d}\sum^{j-1}_{j'=0}f^{(d-1)}_{j'+1}\gamma_\delta^{j'} \\
  &=&\frac{(\gamma_{_W}(1-\gamma_\delta))^d}{d}\sum^{d}_{j=1}\sum^{d-j}_{k=0}(1-\gamma_{_\W})^kf^{(k)}_d\sum^{j-1}_{j'=0}f^{(d-1)}_{j'+1}\gamma_\delta^{j'}.
\end{eqnarray}
where $f^{(d-1)}_{j'+1}=C^{d-1}_{j'+d-1}=C^{j'}_{j'+d-1}=f^{(j')}_{d}$, and we have the following summation exchange
\begin{eqnarray}
   \sum^{d}_{j=1}\sum^{d-j}_{k=0}\sum^{j-1}_{j'=0} 
   \to 
   \sum^{d-1}_{j'=0}\sum^{d}_{j=j'+1}\sum^{d-j}_{k=0}
   \to
   \sum^{d-1}_{j'=0}\sum^{d-j'-1}_{k=0}\sum^{d-k}_{j=j'+1}
\end{eqnarray}
All the terms in the summation are independent of $j$, so the last summation $\sum^{d-k}_{j=j'+1}$ equivalent to multiplying each term by $(d-k-j')$. Hence,
\begin{eqnarray}
 \nonumber  \epsilon^{(d)}&=&\frac{(1-\gamma_\delta)^d}{d}\sum^{d-1}_{j'=0}f^{(j')}_{d}\gamma_\delta^{j'}\gamma_{_\W}^d\sum^{d-j'-1}_{k=0}(d-k-j')f^{(k)}_{d}(1-\gamma_{_\W})^k\\
  &=&\frac{(1-\gamma_{\delta})^d}{d}\sum^{d-1}_{j'=0}(d-j')f^{(j')}_{d}(\gamma_\delta\gamma_{_\W})^{j'}\frac{\gamma^{n-j'}_{_\W}}{d-j'}\sum^{d-j'-1}_{k=0}(d-j'-k)f^{(k)}_{d}(1-\gamma_{_\W})^k
  =I_d(\gamma_\delta , 1-\gamma_{_\W}).
\end{eqnarray}
Here $I_d(x,y)$ is the function defined in Eq. (\ref{eq:Id}). This completes the proof of Theorem $\ref{th:3}$.

\end{widetext}

\newpage %Just because of unusual number of tables stacked at end
\bibliography{230213}% Produces the bibliography via BibTeX.

\end{document}